\documentclass[11pt,a4paper]{article}
\usepackage{jinstpub}

\title{Beam Dynamics in Dielectric Laser Acceleration}

\author[a] {U. Niedermayer}
\author[b] {K. Leedle}
\author[c] {P. Musumeci}
\author[a] {S. A. Schmid}

\affiliation[a] {Technical University Darmstadt, Institute for Accelerator Science and Electromagnetic Fields, Schlossgartenstra\ss e 8, 64289 Darmstadt, Germany}
\affiliation[b] {Ginzton Lab, Stanford University, Stanford, California, USA}
\affiliation[c] {Department of Physics and Astronomy \\
University of California at Los Angeles, Los Angeles, CA, 90095}%

\emailAdd{niedermayer@temf.tu-darmstadt.de}

\date{today}

\abstract{We discuss recent developments and challenges of beam dynamics in Dielectric Laser Acceleration (DLA), for both high and low energy electron beams. Starting from ultra-low emittance nanotip sources the paper follows the beam path of a tentative DLA light source concept. Acceleration in conjuction with focusing is discussed in the framework of Alternating Phase Focusing (APF) and spatial harmonic ponderomotive focusing. The paper concludes with an outlook to the beam dynamics in laser driven nanophotonic undulators, based on tilted DLA grating structures.}

\proceeding{Special Issue on Beam Dynamics Challenges in Advanced and Novel Accelerators\\
  February 2022\\
  }

\begin{document}

\maketitle

\section{Introduction}
The combination of periodic dielectric structures and coherent light allows to reverse the Cherenkov effect and the Smith-Purcell effect~\cite{Smith1953Visible12} in order to attain acceleration of electrons. It was proposed already in 1962~\cite{Lohmann1962ElectronWaves,Shimoda1962ProposalMaser}, shortly after the invention of the laser. The use of dielectric gratings in conjunction with laser light sources has been named Dielectric Laser Acceleration (DLA) after it became a viable approach to accelerate electrons with record gradients. These record gradients are enabled especially by modern ultrashort-pulsed laser systems, mostly in the infrared spectrum, and by nanofabrication techniques for the high damage threshold dielectric materials, as adopted from the semiconductor industry. Due to these high technical demands, the experimental demonstration of electron acceleration in DLA came only in 2013, more than 50 years later than the original proposal~\cite{Peralta2013DemonstrationMicrostructure.,Breuer2013DielectricEffect}. These promising results of gradients, generating only energy spread so far, lead to the funding of the ACHIP collaboration~\cite{2022ACHIPWebsite}, in order to achieve an accelerator attaining MeV energy gain. A summary of DLA research as it stood in 2014 is given in~\cite{England2014DielectricAccelerators}.

Key to the high gradients in DLA is the synchronization of optical near fields to relativistic electrons, expressed by the Wideroe condition 
\begin{equation}
    \lambda_g=m\beta\lambda
    \label{Eq:Wideroe}
\end{equation}
where $\lambda_g$ is the grating period, $\lambda$ is the laser wavelength, and $\beta=v/c$ is the the electron velocity in units of the speed of light. The integer number $m$ represents the spatial harmonic number at which the acceleration takes place. The zeroth harmonic is excluded by means of the Lawson-Woodward theorem~\cite{England2014DielectricAccelerators} as it represents just a plane wave; the first harmonic ($m=1$) is most suitable for acceleration, as it usually has the highest amplitude. Phase synchronous acceleration (fulfilling Eq.~\ref{Eq:Wideroe}) at the first harmonic can be characterized by the synchronous Fourier coefficient
\begin{equation}
    e_1(x,y)=\frac{1}{\lambda_g}\int_{\lambda_g} E_z(x,y,z) e^{2\pi i z/\lambda_g} \mathrm{d}z
    \label{Eq:e1def}
\end{equation}
where $E_z$ is the longitudinal component of the electric field at the laser center frequency in the channel. Often, $E_z$ is normalized to the amplitude of the incident laser field. The transverse dependence of $e_1$ allows also to calculate the the transverse kicks by means of the Panofsky-Wenzel theorem~\cite{Panofsky1956SomeFields,Niedermayer2017BeamScheme}.

The laser systems used in current experiments are mostly 800nm Ti:Sapphire amplifiers providing femtosecond pulses; either directly or followed by an Optical Parametric Amplifier (OPA) to transform into longer infrared wavelengths. Other systems based on 1000~nm sources are also available. Moreover, with modern Holmium / Thulium doped fiber systems, 2~$\mu$m pulses can also be obtained directly. 
Generally, there is a quest for longer wavelength, as this eases the nanophotonic structure fabrication precision requirement.
Working with ultrashort pulses in the femotsecond realm is paramount in DLA, in order obtain high electric fields at low pulse energy. The pulse energy divided by the transverse size defines the fluence, and the acceleration structure material choice is predominantly made by the counteracting figures of merit of high damage threshold fluence and high refractive index, both at given laser wavelength.

Usually the laser pulses are impinging laterally on the structures, with the polarization in the direction of electron beam propagation. Short pulses thus allow interaction with the electron beam only over a short distance. This lack of length scalability can be overcome by pulse front tilt (PFT), which can be obtained from dispersive optical elements, such as diffraction gratings or prisms~\cite{Hebling1996DerivationDispersion}. At ultrarelativistic energy, a 45 degree tilted laser pulse can thus interact arbitrary long with an electron, while the interaction with each DLA structure cell can be arbitrary short. The current record gradient of 850~MeV/m~\cite{Cesar2018High-fieldAccelerator} and record energy gain of 315~keV~\cite{Cesar2018EnhancedLaser} in DLA could be obtained in this way.
Generally, the pulse front tilt angle $\alpha$ must fulfill
\begin{equation}
    \tan \alpha = \frac{1}{\beta}
\end{equation}
in order to remain synchronous with the electron~\cite{Wei2017Dual-gratingLaser,Kozak2018UltrafastNanostructures,Cesar2018High-fieldAccelerator}. This requires a curved pulse front shape, especially for electron acceleration at low energy, where the speed increment is non-negligible. A general derivation of the pulse front shape required for a given acceleration ramp design is given in~\cite{Niedermayer2020ThreedimensionalAccelerators}, where also pulse length minima are discussed when the curved shape is approximated by linear pieces.

Looking towards applications of dielectric laser acceleration, electron diffraction and the generation of light with particular properties are the most catching items, besides the omnipresent goal of creating a TeV collider for elementary particle physics. As such we will look into DLA-type laser driven undulators, which intrinsically are able to exploit the unique properties of DLA-type electron microbunch trains. Morever, the DLA undulators allow creating significatly shorter undulator periods than conventional magnetic undulators, thus reaching the same photon energy at lower electron energy. On the long run, this shall enable us to build table-top lightsources with unprecented parameters~\cite{Rosenzweig2012TheProject}.

This paper is organized along such a light source beamline and the current efforts undertaken towards achieving it. We start with the ultralow emittance electron sources, which make use of the available technology of electron microscopes, to provide a beam suitable for injection into low energy (sub-relativisitic) DLAs.  Then, news from the alternating phase focusing (APF) beam dynamics scheme is presented, especially that the scheme can be implemented in 3D fashion to confine and accelerate the highly dynamical low energy electrons. We also show a higher energy design, as APF can be applied for DLA from the lowest to the highest energies, it only requires individual structures. At high energy, also spatial harmonic focusing can be applied. This scheme gains a lot of flexibility from variable shaping of the laser pulse with a spatial light modulator (SLM) and keeping the acceleration structure fixed and strictly periodic. However, due to the large amount of laser energy required in the non-synchronous harmonics, it is rather inefficient. Last in the chain, a DLA undulator can make use both of synchronous and asynchrounous fields. We outline the ongoing development of tilted DLA grating undulators~\cite{Plettner2008Microstructure-basedLaser,Plettner2008ProposedUndulator,Plettner2009Photonic-basedStructures} towards first experiments in RF-accelerator facilities which provide the required high brightness ultrashort electron bunches.

\section{Ultra-low Emittance Injector}

The sub-400 nm wide accelerator channel and field non-uniformity in dielectric laser accelerators place very strict emittance requirements on the electron injector. Typical acceptances in an APF DLA designed for a 2 micron drive laser require a \textasciitilde10 nm beam waist radius and 1 mrad beam divergence at 30 keV to prevent substantial beam loss during acceleration~\cite{Niedermayer2021DesignChip}. Generating such a 10 pm-rad beam essentially requires the use of a nanometer scale cathode, most commonly implemented via nanotips of various flavors. A variety of nanotip emitters have demonstrated sufficiently low emittance for DLA applications in a standalone configuration, but an additional challenge is to re-focus the beam coming off a tip into a beam that can be injected into a DLA without ruining the emittance. An additional challenge is that most DLA applications require maximum beam current, so generally injection systems cannot rely on filtering to achieve the required emittance. As such, an ultra-low emittance injector for a DLA requires an ultra-low emittance source and low aberration focusing elements to re-image the tip source into the DLA device at 10's to 100's of keV energy. Typical RF photoinjectors and flat cathode electron sources cannot produce < 100nm emittance beams without heavy emittance filtering~\cite{Feist2017UltrafastBeam}. 

There have been two major approaches to producing injectors of sufficiently high brightness. The first approach uses a nanotip cold field or Schottky emitter in an electron microscope column that has been modified for laser access to the cathode~\cite{Feist2017UltrafastBeam,Kozak2018UltrafastNanostructures,Caruso2019HighApplications}. One can then leverage the decades of development that have been put into transmission electron microscopes with aberration corrected lenses. Such systems are capable of producing beams down to 2 nm rms with 6 mrad divergence while preserving about 3 \% of the electrons produced at the cathode~\cite{Houdellier2018DevelopmentSource}, which meets the \textasciitilde10 pm-rad DLA injector requirement for one electron per pulse. Transmission electron microscopes are also available at beam energies up to 300 keV or $\beta = 0.78$, which would reduce the complexity of the sub-relativistic DLA stage compared to a 30 keV or $\beta = 0.33$ injection energy. Moreover, at higher injection energy, the aperture can be chosen wider, which respectively eases the emittance requirement. Laser triggered cold-field emission style TEMs tend to require more frequent flashing of the tip to remove contaminants and have larger photoemission fluctuations  than comparable laser triggered Schottky emitter TEMs ~\cite{Feist2017UltrafastBeam, Houdellier2018DevelopmentSource}. 

\begin{figure}
    \centering
    \includegraphics[width=0.8\textwidth]{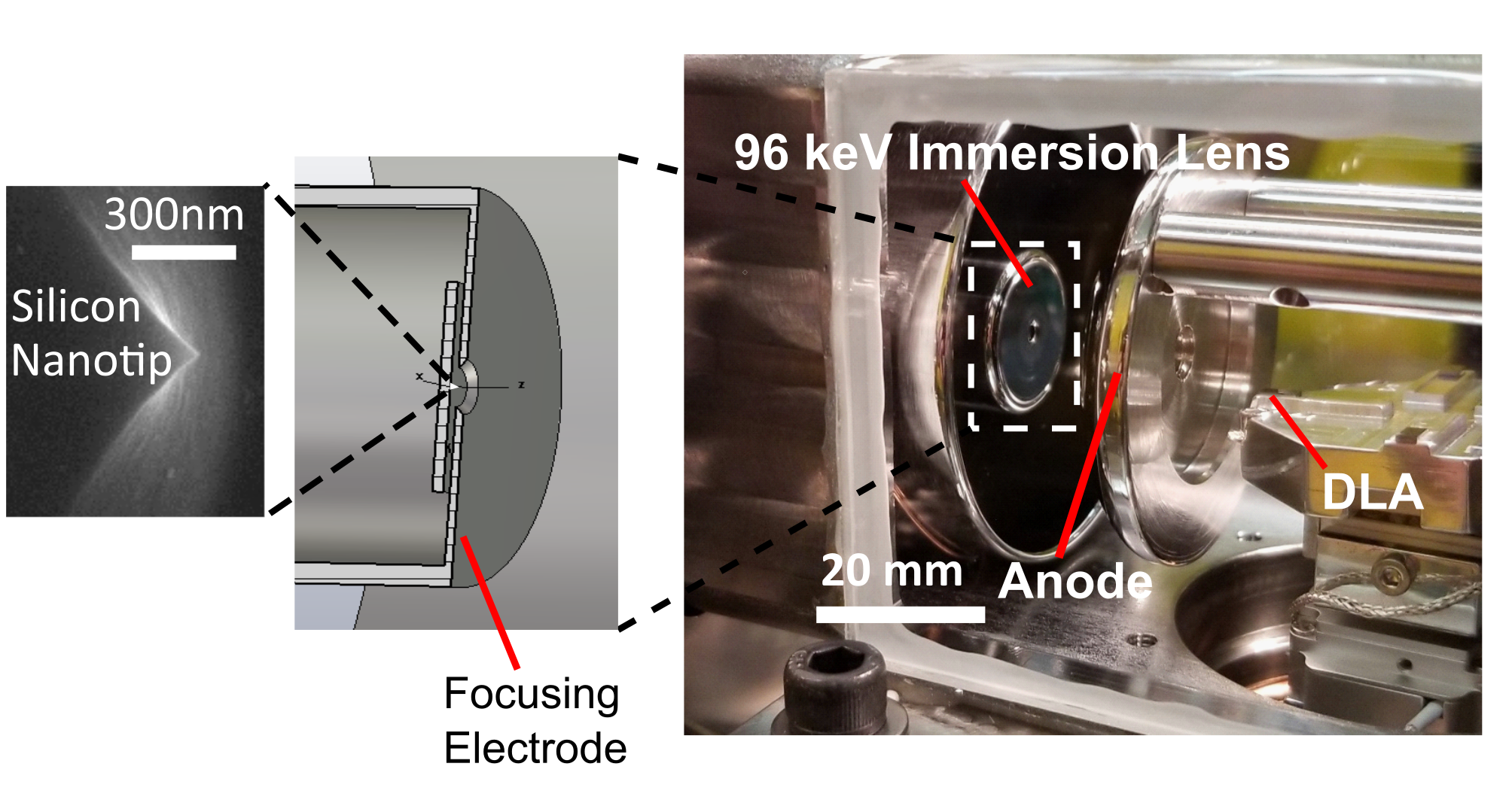}
    \caption{Schematic of the 96 keV Glassbox Immersion Lens, which uses a single electrode to electrostatically focus the beam from a silicon nanotip into a DLA accelerator at a 25~mm working distance.}
    \label{fig:GlassboxSch}
\end{figure}

A major disadvantage of using modified transmission electron microscopes as the injector for a DLA is their large size and cost. A typical 200 keV system will have roughly a 1 meter distance from the electron source to the DLA. Efforts have been made to produce a a more compact electron injector by using an electro-static immersion lens built into the electron source region to directly focus the beam coming off of a nanotip source~\cite{Hirano2020AAccelerator}. This allows the nanotip emitter to be re-imaged into a DLA device in less than 25~mm distance at 96 keV beam energy as shown in Fig.~\ref{fig:GlassboxSch}~\cite{Leedle2022HighSources}. Due to the relatively high beam divergence off the tip feeding directly into the immersion lens, this arrangement is more prone to aberrations than a system with focusing elements outside the electron acceleration region. As such, a 0.4~mrad divergence beam out of an immersion lens source will typically be focusable to 400 nm rms without emittance filtering, resulting in a normalized emittance of ~100 pm-rad, increasing to 300~pm-rad at 30 electrons per shot, which is better than the unfiltered emittance from a CW laser triggered Schottky emitter TEM~\cite{Feist2017UltrafastBeam}. The immersion lens reduces the DC acceleration gradient at the tip, increasing the effects of space-charge compared to other source designs. The initial immersion lens experiments were also performed using five-photon photoemission from silicon nanotips to suppress background emission, which results in a larger space charge energy spread than necessary~\cite{Schenk2010Strong-fieldTips}.

Moving forward, there are possibilities of increasing the acceleration gradients in DC biased electron sources through the use of novel electrode materials and preparation~\cite{LePimpec2010VacuumSources,Leedle2022HighSources} to 20 to 70 kV/mm or greater gradients. These higher gradients, combined with the local field enhancement at a nanotip emitter to just short of the field emission threshold, will result in higher brightness and lower emittance beams to be focused by downstream optics. Triggering the cathode with laser pulses matched to its work function also helps minimize the beam energy spread and reduce chromatic aberrations~\cite{Schenk2010Strong-fieldTips}. Beam energy spread is determined by the net effects of space-charge and excess cathode trigger photon energy, and can range from 0.6 eV FWHM for low charge single-photon excitation to over 5 eV FWHM for 100 electrons per shot~\cite{Feist2017UltrafastBeam}. This excess energy spread increased the electron bunch duration from a minimum of 200 fs to over 1 ps FWHM at high charge in a TEM~\cite{Feist2017UltrafastBeam}. The majority of the space charge induced energy spread occurs within a few microns of the emitter, emphasizing the importance of having a maximum acceleration field at the emitter~\cite{Bach2019CoulombEmitters}.

\section{Alternating Phase Focusing DLA}
\subsection{Principle and Nanophotonic Structures}
The advantage of high gradient in DLA comes with the drawback of non-uniform driving optical nearfields across the beam channel. The electron beam, which usually fills the entire channel, is therefore defocused. The defocusing is resonant, i.e. happening in each cell, and so strong that it cannot be overcome by external quadrupole or solenoid magnets~\cite{Ody2017FlatAccelerators}. The nearfields themselves can however be arranged in a way that focusing is obtained in at least one spatial dimension. Introducing phase jumps allows to alternate this direction along the acceleration channel and gives rise to the Alternating Phase Focusing (APF) method~\cite{Niedermayer2018Alternating-PhaseAcceleration}.  
The first proposal of APF for DLA included only a twodimensional scheme (longitudinal and horizontal) and the assumption that the vertical transverse direction can be made invariant. This assumption is hard to hold in practice, since it would require structure dimensions (i.e. pillar heights) that are larger than the laser spot. The laser spot size, however, needs to be large for a sufficiently flat Gaussian top. Moreover, if the vertical dimension is considered invariant, the beam has to be confined by a single external quadrupole magnet~\cite{Niedermayer2018Alternating-PhaseAcceleration, Niedermayer2019ChallengesAcceleration}, which requires perfect alignment with the sub-micron acceleration channel. While these challenges can be assessed with symmetric flat double-grating structures with extremely large aspect ratio (channel height / channel width), it is rather unfeasible with pillar structures where the aspect ratio is limited to about 10-20. For these reasons, the  remaining vertical forces render the silicon pillar structures suitable only for limited interaction length.

\begin{figure}[b]
    \centering
    \includegraphics[width=0.8\textwidth]{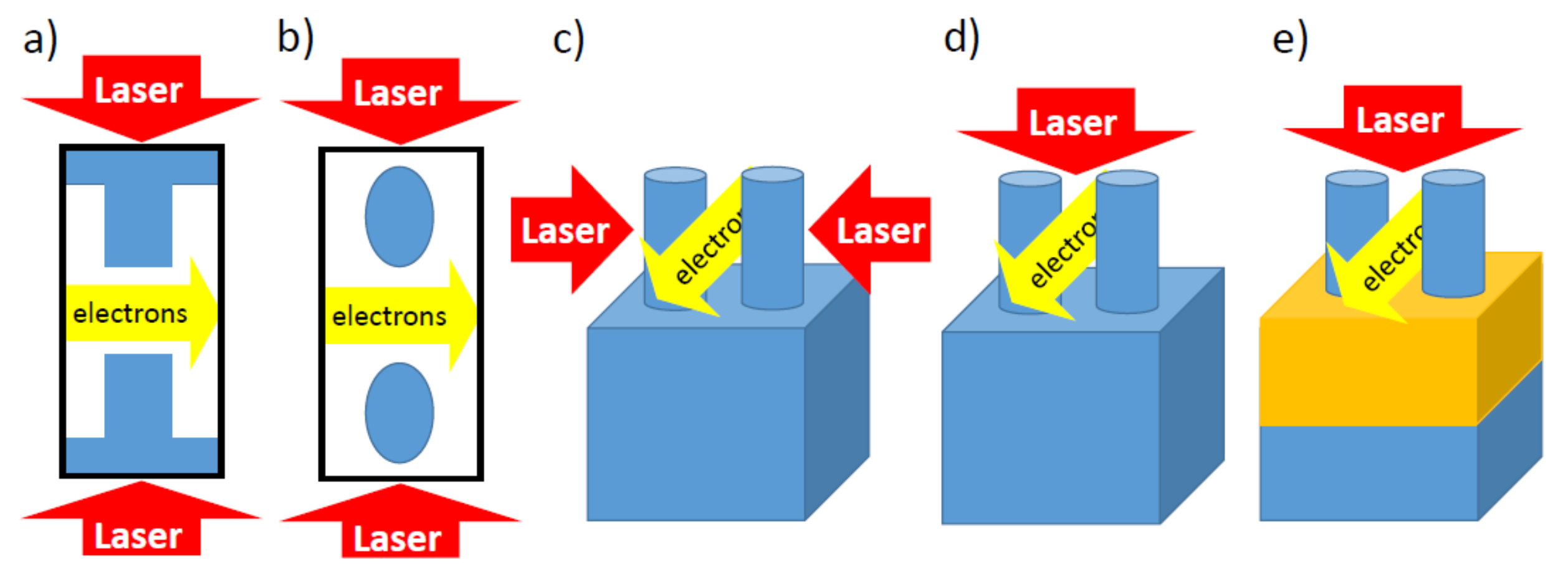}
    \caption{Low energy DLA cells: Pillar (a) and grating (b) structures in 2D approximation and pillar structures as they are in reality. The structures (c) and (d) suffer of fluctuations of the structure constant along the vertical coordinate. The structure in (e) exploits this effect to create focusing in the vertical direction by using two materials with high (blue) and low (orange) refractive index.}
    \label{fig:DLAcell2}
\end{figure}

Different types of DLA structures (single cells) are described in Fig.~\ref{fig:DLAcell2} for low energy (sub-relativistic) and in Fig.~\ref{fig:DLAcell3} for high energy (relativistic).
The usual materials are silicon or fused silica (SiO\textsubscript{2}), which can be nanofabricated by established techniques from the semiconductor industry. The vertical confinement problem can be solved by using a two-material wafer with high refractive index contrast as shown in Fig.~\ref{fig:DLAcell2} (e). This gives rise to the threedimensional APF for DLA scheme~\cite{Niedermayer2020ThreedimensionalAccelerators} which finally makes DLA fully length scalable.

From Maxwell's equations and the phase synchronicity condition Eq.~\ref{Eq:Wideroe} one can derive the synchronous longitudinal field as  
\begin{equation}
    e_1(x,y)=e_{10}\cosh(ik_x x)\cosh(ik_y y), 
    \label{Eq:e1}
\end{equation}
where $e_{10}$ is usually referred to as the structure constant. It needs to be determined for each structure individually (by numerical techniques) and takes values between 0.05 and 1.2 in practical cases.
The propagation constants fulfill the dispersion relation
\begin{equation}
 k_x^2+k_y^2 = -\frac{\omega^2}{\beta^2\gamma^2 c^2},     
\end{equation}
where $\omega=2\pi c/\lambda$ is the laser angular frequency and $\beta,\gamma$ are the relativistic velocity and mass factors. 
Note that the longitudinal field Eq.~\ref{Eq:e1} suffices to describe the entire threedimensional kick on a particle in the DLA cell, since the transverse kick can be calculated by the Panofsky-Wenzel theorem~\cite{Niedermayer2017BeamScheme}. The horizontal propagation constant $k_y$ is always purely imaginary, that is an inherent consequence of DLA being a nearfield acceleration scheme. The vertical propagation constant $k_x$ can however be either purely imaginary or purely real. If $k_x$ is purely imaginary (as $k_y$), we call the focusing scheme 'in-phase', since both transverse directions are simultaneously focused, while the longitudinal direction is defocused, and vice versa. Oppositely, if $k_x$ is real valued, the vertical direction (x) focuses simultaneously with the longitudinal direction, exactly when the horizontal direction (y) is defocusing and vice versa; the scheme is thus called 'counter-phase'. 

Technically, counter-phase structures are much easier to realize than in-phase structures, since in-phase structures require passing the electron beam through a vertical region of low refractive index, i.e. two high-refractive index layers above and below the electron beam would be required~\cite{Niedermayer2020ThreedimensionalAccelerators}. Particularly, counter-phase structures can be implemented on Silicon on Insulator (SOI) wafers, which are commercially available and their processing is standard in nanophotonics. Moreover, the pillars on SOI can be grounded to remove lost electrons by individual lateral traces, without deteriorating the structure bandwidth~\cite{Niedermayer2020ThreedimensionalAccelerators,Niedermayer2021DesignChip}.

\begin{figure}[t]
    \centering
        \includegraphics[width=0.6\textwidth]{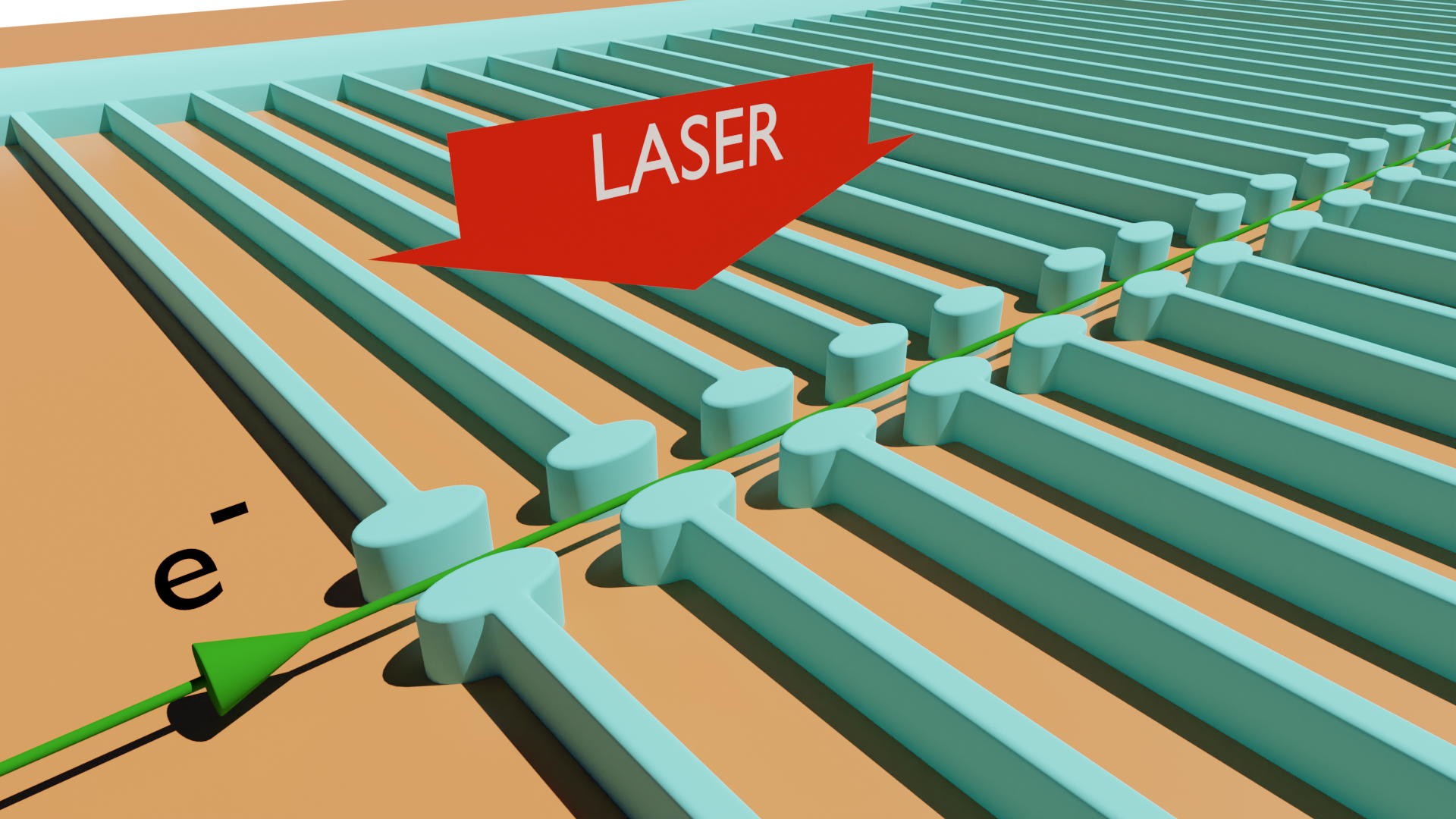}
    \includegraphics[width=0.3\textwidth]{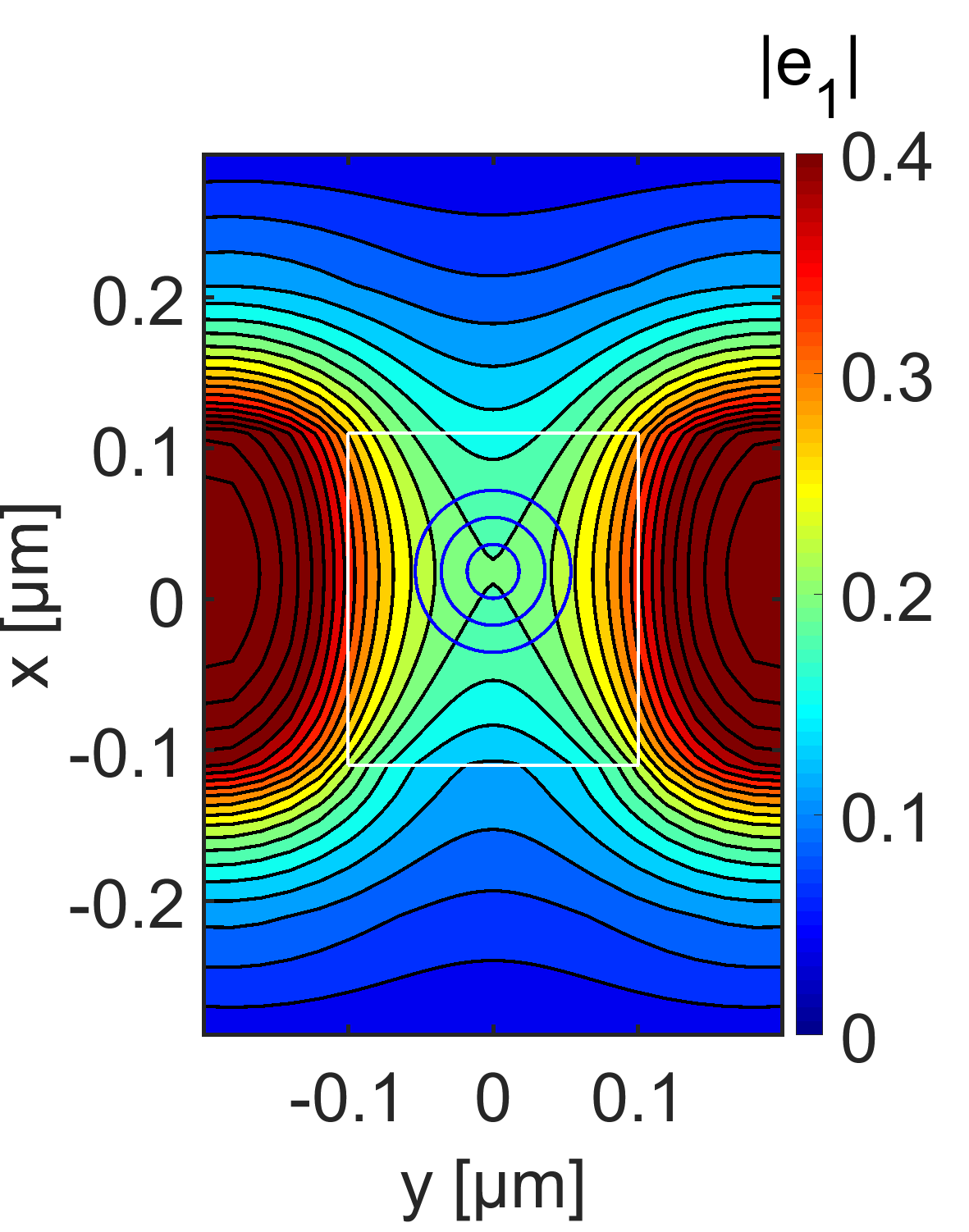}
        \includegraphics[width=0.85\textwidth]{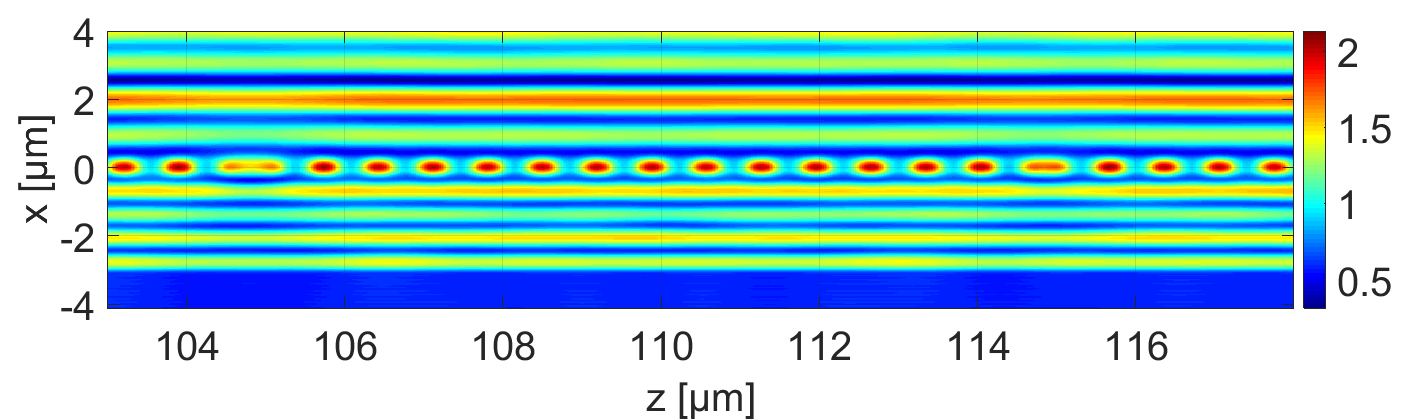}
    \caption{Top left: Accelerator structures on SOI chip. Top right: Distribution of normalized $e_1(x,y)$. The blue circles indicate 1,2,3 $\sigma$ of the electron beam, the white box indicates the channel. Bottom: normalized longitudinal electric field distribution in a longitudinal cut. The laser is mostly reflected at the substrate-to-oxide interface. The elongated field maxima deviating from the periodicity are the APF phase jumps.  }
    \label{fig:fields}
\end{figure}
\begin{figure}[t]
    \centering
    \includegraphics[width=1.0\textwidth]{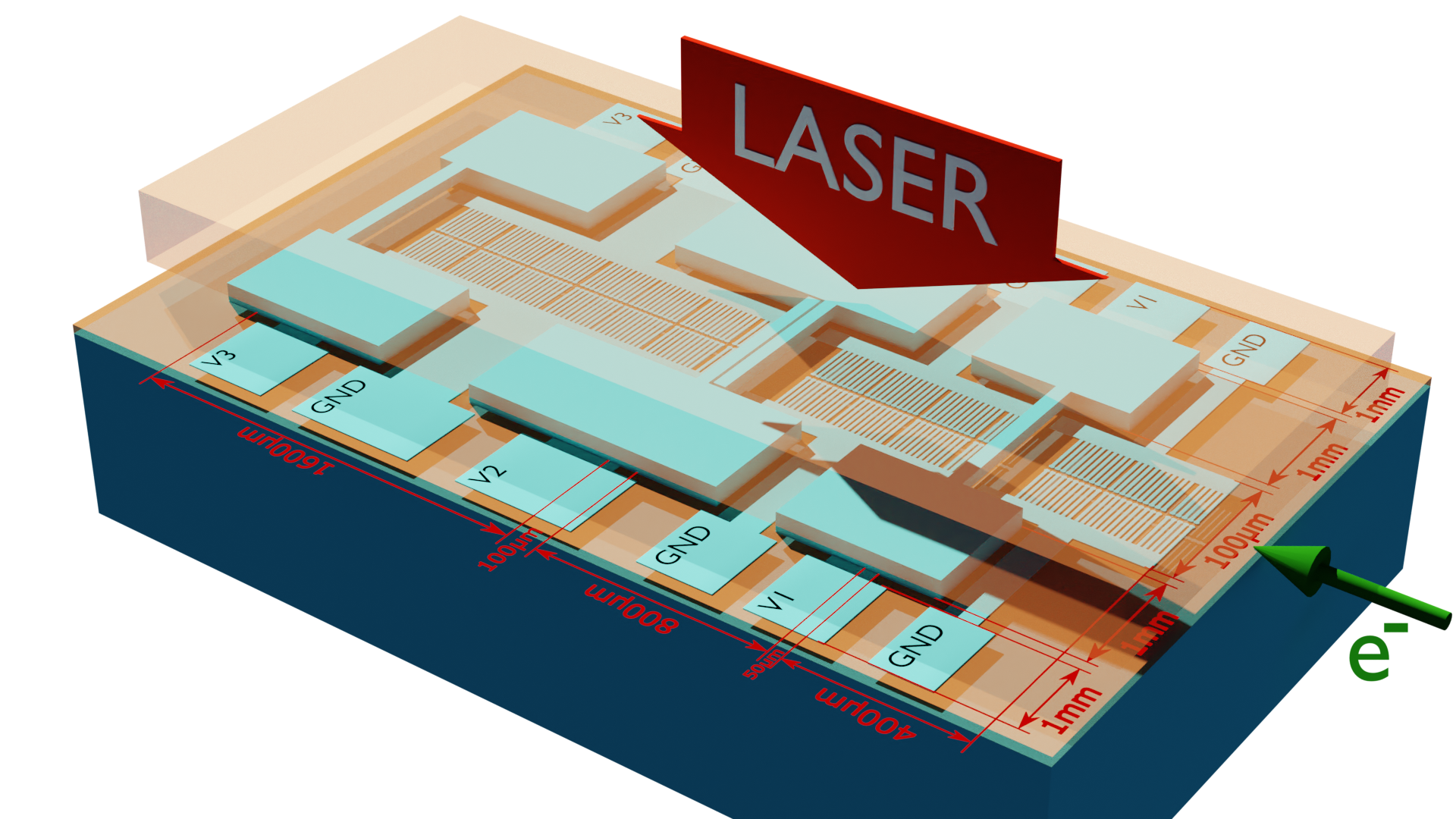}
    \caption{Mock-up design of a completely integrated multi-stage accelerator on a SOI chip. The drawing is not to scale, however the attached dimensions are roughly realistic. }
    \label{fig:AcceleratorChip}
\end{figure}

The 3D APF on SOI structures are shown in Fig.~\ref{fig:fields}, along with a plot of $e_1(x,y)$, exhibiting the quadrupolar shape inherent to the counter-phase scheme. Furtheremore it can be seen that a field enhancement is created by the reflection from the oxide to substrate interface. A complete threedimensional accelerator design on a commercial SOI wafer, doubling the electron energy is shown in~\cite{Niedermayer2021DesignChip}, alongside with a complete start to end simulation, first in DLAtrack6D~\cite{Niedermayer2017BeamScheme} and then a full simulation in CST Particle Studio~\cite{2021CSTSuite}. A major finding of this extensive simulation study was that the vertical phase gradient $\partial_x\arg e_1(x,y)$ needs to be controlled, as it exerts a constant vertical force similar to a homogeneous magnetic field that would drive the beam outside the channel. Controlling the phase gradient can be done by shaping the pillars or by adjusting the ground connection trace longitudinal position. Finally, the result is that a 10pm transverse geometric emittance beam can be accelerated from 26.5keV to 53keV over a length of less than 0.4mm with a throughput of roughly 50\%.

As shown in Fig.~\ref{fig:AcceleratorChip}, multiple acceleration stages can be arranged on a single SOI chip. Each stage roughly doubles the energy and is characterized by the laser pulse front tilt angle, corresponding to an 'average' beam velocity in the stage (See~\cite{Niedermayer2020ThreedimensionalAccelerators} Supplementary material for the optimal constant tilt angle within a stage). 
Between the stages, a vertical adjustment of the beam position can be done by electrostatic steerers, which use the substrate and another silicon on glass wafer, attached from the top as two plates of a deflecting capacitor. The contacting can be done on the device layer of the SOI wafer. Due to the small distance of the plates, voltages of only about 30V are sufficient to obtain sufficiently large deflections to counteract accumulated deflection errors over hundreds of periods.

\subsection{Low Energy Applications and Experiments}
\begin{figure}[bth]
    \centering
    \includegraphics[width=0.61\textwidth]{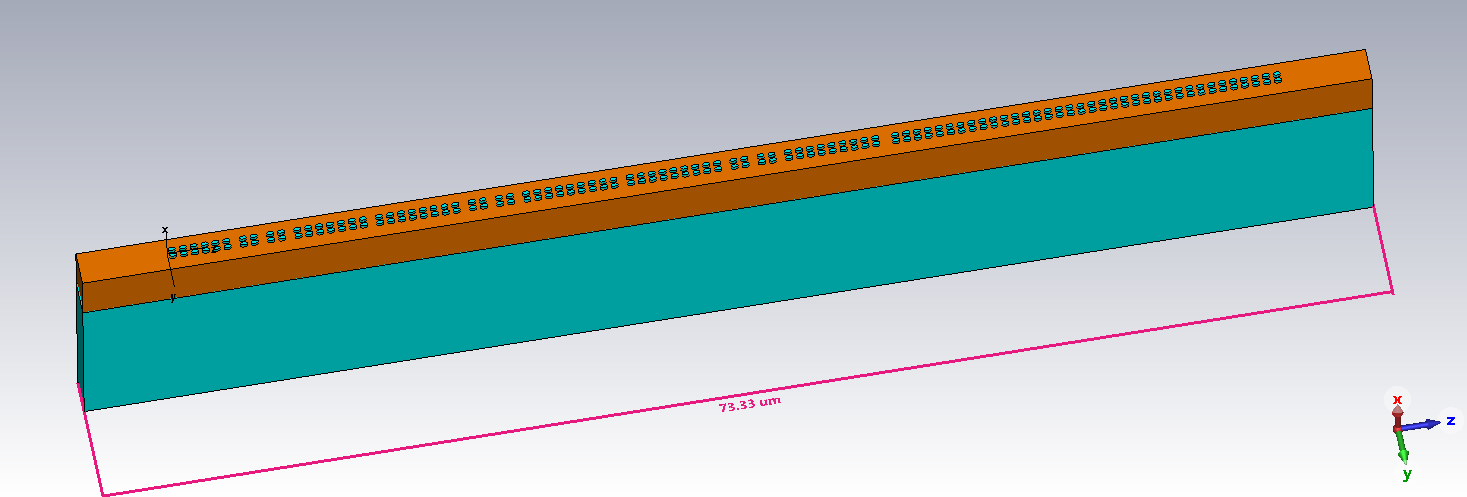}
        \includegraphics[width=0.38\textwidth]{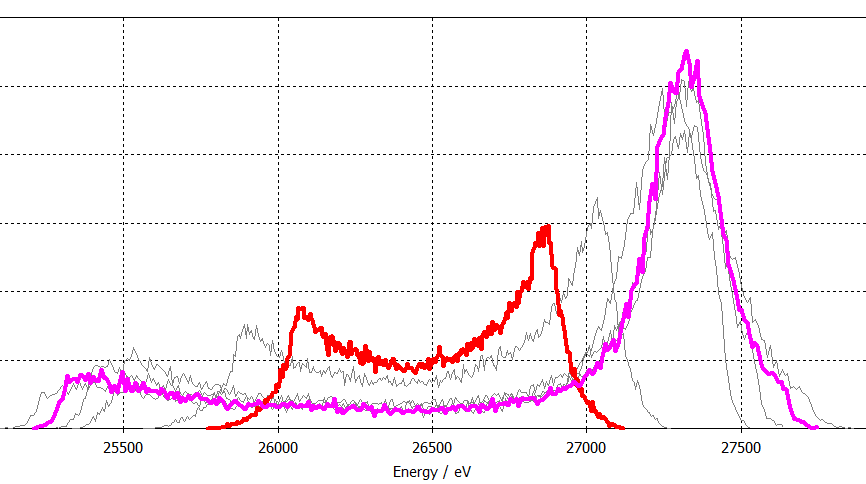}
    \caption{Left: 3D APF buncher realized by three modulate-transport-demodulate stages (consisting of 4 segments each) and accelerator (last long segment). The laser is impinging from the top on the SOI wafer. Right: Energy spectra from full 3D simulation for different laser amplitudes (Red: 136~MV/m Violet: 400~MV/m)}
    \label{fig:ErlangenFancyBuncher}
\end{figure}

At low energy, acceleration gradients are not that critical, since an accelerator chip will only be of the size of a thumbnail to reach relativistic velocity, which we define as 1~MeV electron energy. Therefore, gradient can be sacrificed to some extend for flexibility and improved beam confinement. The first sacrifice is the utilization of materials which are DC-conductive and have a high refractive index, but suffer a significantly lower laser damage threshold. The best example of such is silicon, which also allows us to use the wide range of semiconductor fabrication tools. By etching the pillars by electron beam lithography and the 'mesa' by photo lithography, several low energy electron manipulation devices, well known in the accelerator toolbox, were created on a chip. These are ballistic bunchers~\cite{Black2019NetAccelerator,Schonenberger2019GenerationAcceleration}, APF single cells and channels~\cite{Black2019Laser-DrivenMicrostructures,Shiloh2021ElectronAcceleration}, and the first demonstration of low energy spread bunching and coherent acceleration in DLA~\cite{Niedermayer2021LowNanostructures}. Yet all these devices suffer from lack of real length scalability due to a 2D design with insufficient pillar height. Moreover, the coherent acceleration experiment did not attain the energy spread as low as predicted by 2D simulations. The reason for this is the fluctuation of the structure constant $e_1$ as function of the vertical coordinate in conjunction with the beam being unconfined vertically. By building a 3D APF multistage buncher, energy spreads as low as predicted in the 2D simulations have been demonstrated in full 3D simulations, and should thus be achieved in experiments soon. A design and full 3D field and particle simulation of such a multi-stage 3D APF buncher and accelerator is shown in Fig.~\ref{fig:ErlangenFancyBuncher} 

\subsection{High Energy High Gradient Acceleration}
\begin{figure}[bh]
    \centering
    \includegraphics[width=0.8\textwidth]{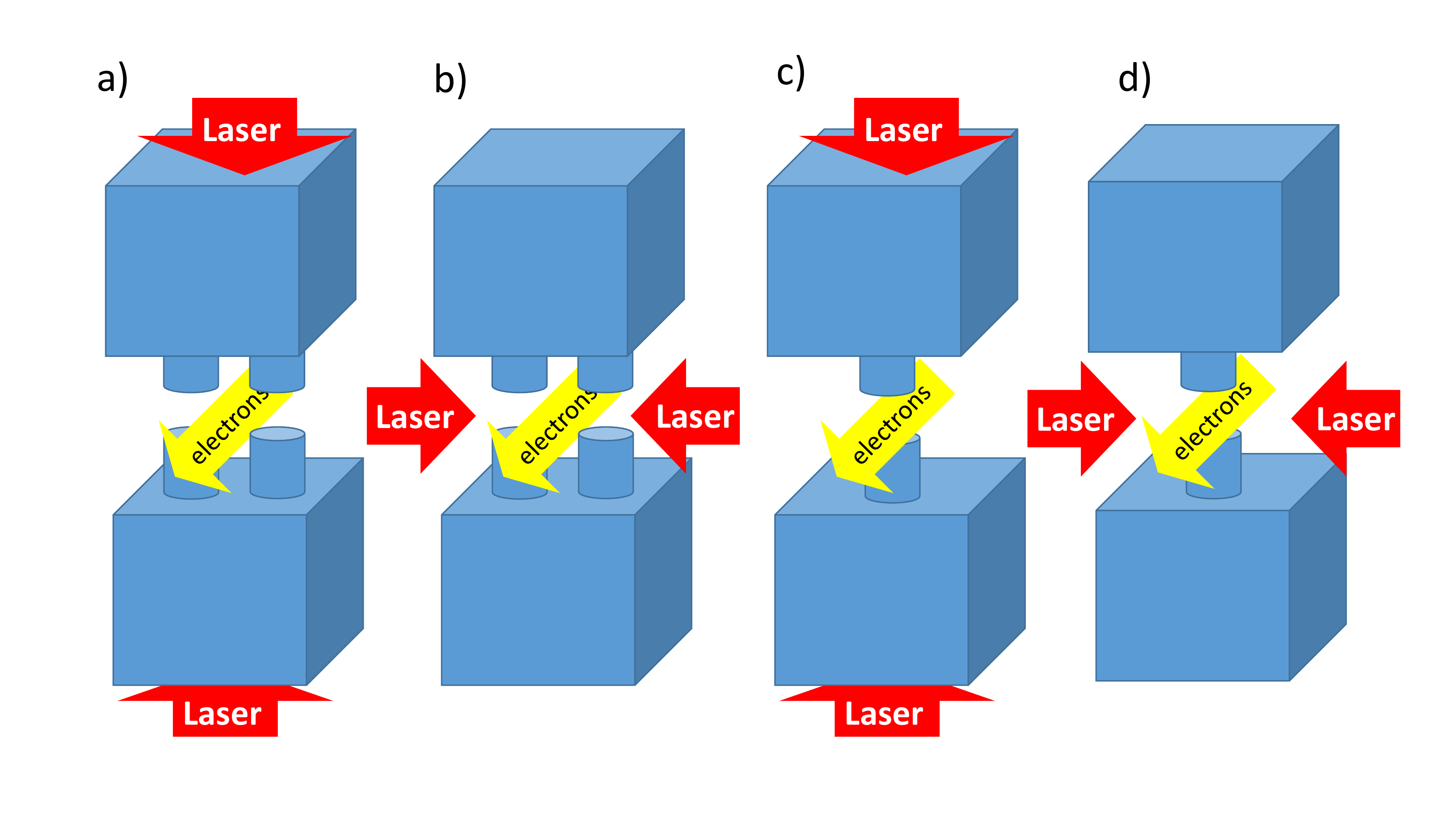}
    \caption{Symmetric high energy high gradient DLA single cells. All options create a counter-phase scheme quadrupole disribution at the electron beam.}
    \label{fig:DLAcell3}
\end{figure}

\begin{figure}[h]
    \centering
    \includegraphics[width=0.9\textwidth]{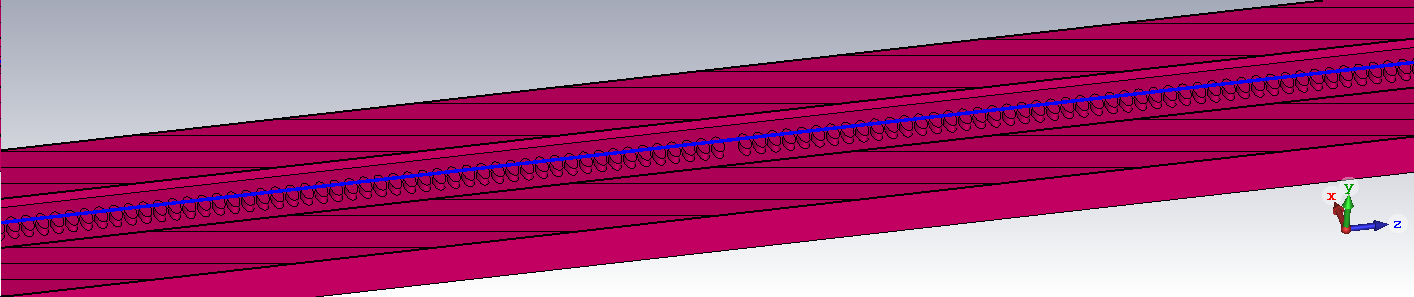}
        \includegraphics[width=0.95\textwidth]{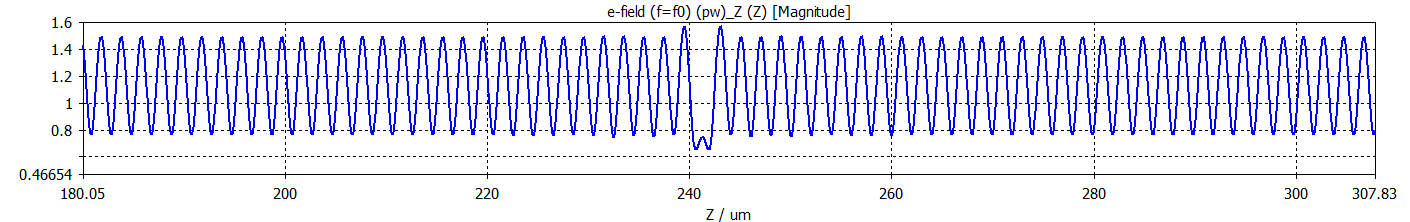}
                \includegraphics[width=1.0\textwidth]{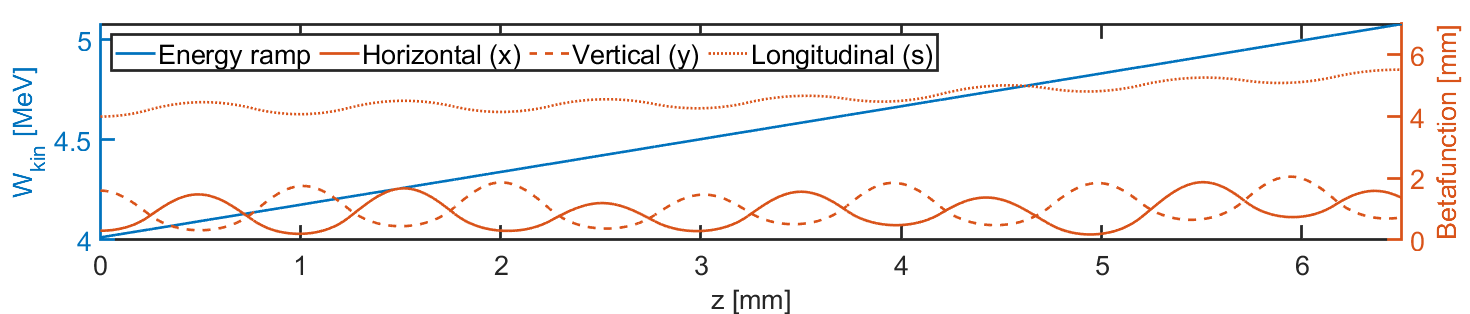}
    \caption{3D APF counter-phase structure for relativistic acceleration. Top: half of the structure (The laser comes from top and bottom as in Fig.~\ref{fig:DLAcell3} (c), the blue line is the beam axis), center: normalized electric field on axis, bottom: energy ramp and betafunctions of the six-dimensional confinement.}
    \label{fig:highE}
\end{figure}

In order to exploit the unique features of DLA for a high energy accelerator, a high damage threshold material has to be used. A list of such materials is provided in~\cite{Soong2012LaserAccelerators}. A particular material which was used to obtain the record gradients the experiments is Fused Silica (SiO\textsubscript{2})~\cite{Peralta2013DemonstrationMicrostructure.,Wootton2016DemonstrationPulses,Cesar2018High-fieldAccelerator}. By bonding two SiO\textsubscript{2} gratings together a symmetric structure is obtained, however, in order to obtain the symmetric fields in the channel also the laser illumination must be symmetric. Theoretically, a Bragg mirror could also be used here, however its fabrication using layers of SiO\textsubscript{2} and vacuum is technically challenging. Moreover, the bonded grating structures are essentially 2D, i.e. the laser spot is smaller than the large aperture dimension. This leads to the small focusing strength as discussed above as $k_x=0$ and $k_y=i\omega/(\beta\gamma c)\rightarrow0$ for $\beta\rightarrow 1$. Strong improvement comes from applying 3D APF in the counter-phase scheme. Structures for this are depicted in Fig.~\ref{fig:DLAcell3}. Note that for highly relativistic velocities the in-phase scheme is practically impossible as Eq.~\ref{Eq:e1} implies that in this case $e_1(x,y)$ should be constant, and matching with the boundary conditions implies that it must be close to zero. 
We show an example of casting the counter-phase structures in Fig.~\ref{fig:DLAcell3} (c) and (d) into an accelerator gaining 1~MeV at 4~MeV injection energy. The design relies on etching a trench into a SiO\textsubscript{2} slab and leaving out a pillar row with APF phase jumps, see Fig.~\ref{fig:highE}. By direct bonding of two such slabs, 3D APF structures of a single material, as shown in Fig.~\ref{fig:DLAcell3} (c) and (d), are obtained.
At a synchronous phase 30 degrees off crest and 500~MV/m incident laser field from both sides, about 3000 periods (6~mm total length) are required to obtain 1~MeV energy gain. Figure~\ref{fig:highE} shows the structure, the electric field, and the betafunctions of a designed lattice containing 7 focusing periods. This structure, or respectively lattice, is not yet optimized. The parameters, including the 800~nm aperture, were chosen rather arbitrarily. A preliminary DLAtrack6D simulation shows that an energy gain of 1~MeV with a throughput of about 70\% can be obtained with about 0.08~nm~rad normalized emittance and 0.08~fs FWHM bunchlength.

\section{Spatial Harmonic Focusing}
One important assumption in the APF discussion is that particles mainly interact with the resonant electromagnetic waves as they propagate in the DLA structure. That is the effect of waves propagating with phase velocity different than the velocity of the electrons averages out and does not play a significant role in the dynamics~\cite{Niedermayer2017BeamScheme}. For relativistic DLAs where the velocity of the particles and of the waves inside the structure are all close to the speed of light (and their differences tends to zero), this approximation needs to be carefully reviewed. If the difference in phase velocity is not very large, in fact, even the non resonant harmonics in the field (which can be always modeled as a sum of spatial harmonics) will contribute to the dynamics. 

An interesting alternative scheme for focusing has been proposed by Naranjo et al.~\cite{Naranjo2012StableHarmonics} and borrows from the concept of second order focusing in conventional RF structures~\cite{Reiche1997ExperimentalAccelerator}. In that paper the authors considered a structure with a super-period length $\lambda_S=2\pi / \delta_k$ added to the accelerating wave. Then, using Floquet’s theorem, the field can be decomposed into spatial harmonics. Due to the different phase velocities of these harmonics, only one of these can be resonant (i.e. maintain a nearly constant phase) with the electrons, while all the other ones will wash out over the beam as it propagates in the DLA. The alternation of focusing and defocusing phases would have a net focusing effect which can be used to counteract the resonant defocusing force and maintain the beam confined in the accelerating channel. 

\subsection{Ponderomotive focusing}
To simplify and generalize the discussion one can imagine the electrons to be moving in the field of two waves of electric field amplitude $E_0$ and $E_1$ respectively, one with velocity $\beta_r$ phase synchronous with the beam (which provides stable longitudinal acceleration, but with a net defocusing effect) and the other one at a slower velocity $\beta_1 \cong \beta_r-\delta_k/k$ where $k=2\pi/\lambda_g$ is the wavenumber of the DLA structure.  In a 2D perfect slab-geometry, we can neglect the motion in the long structure dimension and close to the axis the equations of motion for the electron deviation from the resonant energy $\delta\gamma$ and for its normalized transverse velocity $y'$ can be written as
\begin{equation}
    \begin{split}
    \frac{\partial \delta \gamma}{\partial z} &= -\frac{qE_0}{m_0c^2} \sin{\phi} - \frac{qE_1}{m_0c^2} \sin{(\phi + \delta_k z)} \\
    \frac{\partial y'}{\partial z} &= \frac{qE_0}{m_0c^2} \frac{ky}{\gamma^3 \beta^2}\cos{\phi}+\frac{qE_1}{m_0c^2}\frac{ky}{\gamma\beta} (1- \beta \beta_1)\cos{(\phi+\delta_k z)},
    \label{two_wave_DLA}
    \end{split}
\end{equation}
where $m_0c^2/q=511$~kV is the rest energy equivalent and $\phi$ is the phase within one DLA cell.
Analyzing the solutions of the equation for transverse motion into a slowly varying secular component and a fast oscillation, we can rewrite for the slow drift motion 
\begin{equation}
    \frac{\partial y'}{\partial z} = \frac{qE_0}{m_0c^2} \frac{ky}{\gamma^3 \beta^2}\cos{\phi}-\left[\frac{E_1}{m_0c^2}\frac{k}{\gamma\beta} (1- \beta \beta_1)\right]^2\frac{y}{2\delta_k^2}
\end{equation}
and noting that the coefficient second term is negative for all phases, retrieve the ponderomotive focusing effect~\cite{Szczepkowicz2017ApplicationElectrons}. The main drawback of the ponderomotive focusing scheme (compared to the APF scheme discussed above) is the significant need for power in the non-resonant harmonic $E_1$ to compensate the strong resonant defocusing, so that the laser is not efficiently used to accelerate the particles (i.e. $E_0$ is relatively small). 
Interestingly, in 2D APF focusing schemes the focusing term scales with the energy as $1/\gamma^3$, here it scales as $1/\gamma^2$ and for the 3D APF scheme it scales as $1/\gamma$~\cite{Niedermayer2020ThreedimensionalAccelerators}, which eventually dominates over the resonant defocusing scaling as $1/\gamma^3$. Thus, spatial harmonic focusing provides a matched (average) beta function proportional to the beam energy, perfectly compensating the adiabatic geometric emittance decrease to provide a constant spot size along the accelerator. 
 
\subsection{Soft tuning of DLA parameters}
\begin{figure}[th]
    \centering
    \includegraphics[width=1.0\textwidth]{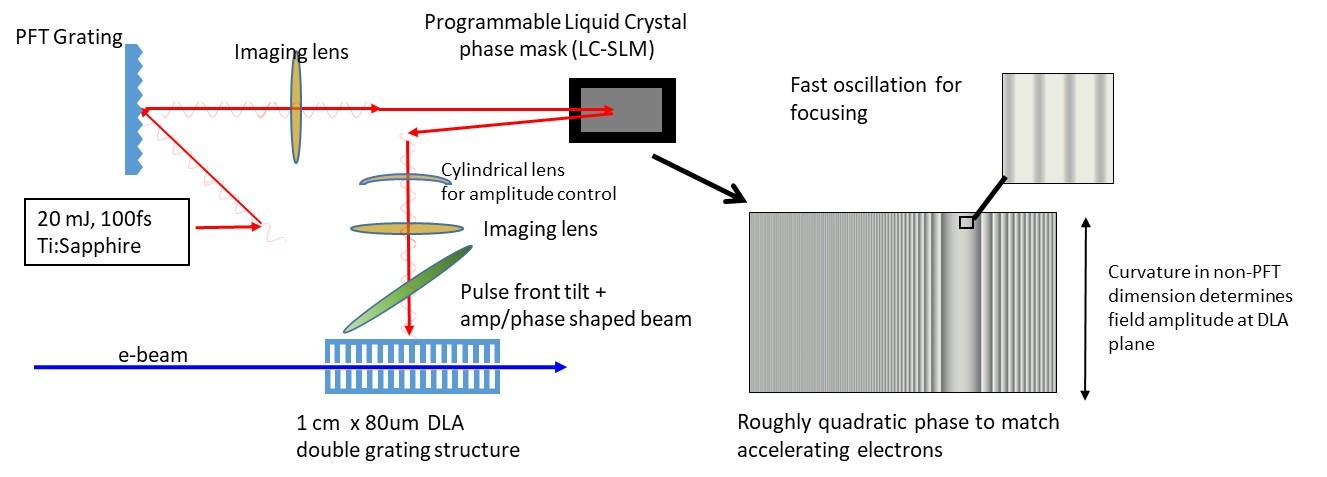}
    \caption{Schematics of a soft tuning scheme combining pulse-front-tilt and liquid-crystal-mask. Imaging lenses ensure that the correct PFT angle is obtained. The cylindrical lenses can be used in combination with curvature on the non-PFT dimension to impart amplitude modulation in addition to the phase modulation at the DLA plane.}
    \label{fig:soft_tuning}
\end{figure}
\begin{figure}[bh]
    \centering
    \includegraphics[width=1.0\textwidth]{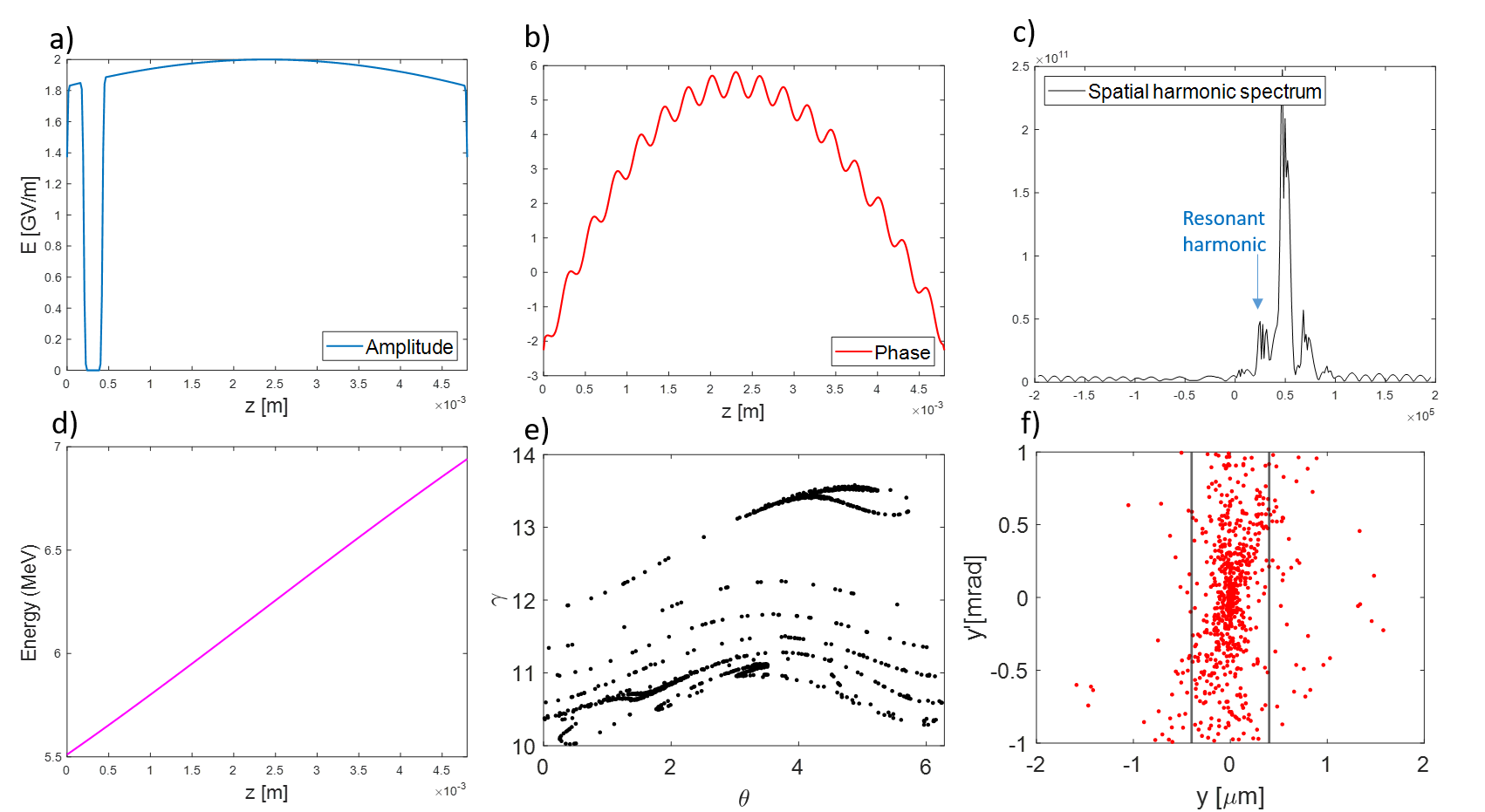}
    \caption{SHARD simulation: a) Electric field amplitude; b) Electric field phase; c) Spatial harmonic spectrum; d) Resonant energy along DLA; e) Longitudinal phase space; f) transverse phase space at the output of the DLA. The simulation assumes a 0.1 \% energy spread, 5.5 MeV eneergy beam with normalized emittance 0.1 nm focused to a 0.25 $\mu$m spot at the entrance of the DLA. More than 20 \% of the particles are captured and accelerated to the final energy.}
    \label{fig:shard}
\end{figure}

The original plan proposed to hard-wire spatial harmonics into the structure to obtain the ponderomotive focusing effect. In practice, one can also simply modulate the drive laser phase, effectively introducing spatial harmonics into a generic, strictly periodic grating, see Fig.~\ref{fig:soft_tuning}. This is due to the low Q-factor of the structures used for DLA, so that the fields in the electron beam channel are actually faithful reproductions of the illuminating laser pulses. Therefore, dynamically controlling phase and amplitude of the drive laser actually offers an interesting alternative to soften the tight tolerance requirements on structure fabrication and enable tuning of the accelerator characteristics without the need to modify/manufacture delicate and expensive dielectric structures \cite{Crisp2021ResonantAccelerator}.

While it is likely that in the future phase and amplitude control of the drive pulses will be implemented using on-chip laser manipulation \cite{Hughes2018On-ChipAccelerators, Sapra2020On-chipAccelerator}, the first exploratory research can be carried out using free space coupling combining pulse front tilt illumination with modern technologies readily available for nearly arbitrary shaping of laser fields in the transverse plane.

Pulse front tilt can be easily coupled with standard methods for spatial light manipulation such as digital micromirror devices or liquid crystal masks \cite{Cesar2018OpticalAccelerator}. Exploiting the 2D nature of these devices, they can be used to apply not only arbitrary phase, but also arbitrary amplitude masks to the transverse profile of the laser which gets converted by the pulse front tilt illumination into temporal modulation seen by the electrons. As masks can be changed essentially online, at very high repetition rates (up to KHz), such system will allow to fine tune the DLA output beam parameters online, with direct guidance from beam diagnostics. In the experimental phase this will also allow testing of various beam dynamics control approaches, including alternate phase focusing, ponderomotive focusing or anything else in between. A new code has been developed to self-consistently calculate the interaction of relativistic particles with different phase velocities spatial harmonics \cite{Ody2021SHarD:Expansion} (see Fig. \ref{fig:shard}). 

In fact, with a full control over amplitude and phase seen by the particles the question becomes how to best optimize the drive pulse. At this purpose, it is envisioned adopting machine learning algorithms to optimize the 2D mask which gets applied to the laser drive pulse by the liquid crystal modulator~\cite{Edelen2020MachineSystems}.

\section{DLA Undulator}
        Similar to a conventional magnetic undulator, a DLA undulator needs to provide an oscillatory deflection force as well as transversal confinement to achieve stable beam transport and scalable radiation emission.  On the long run, a DLA based radiation source would use beams provided by a DLA accelerator. However, closer perspectives to experiments favor using advanced RF accelerators which can also provide single digit femtosecond bunches at high brightness. The ARES accelerator~\cite{Cankaya2021TemporalRegime} at SINBAD/DESY provides such beam parameters suitable to be injected into DLA undulators. Thus, we adapt our design study on the 107~MeV electron beam of ARES. 
        
    \subsection{Tilted Grating Design}
\begin{figure}
        	\begin{minipage}{0.49\textwidth}
        	    \vspace{0pt}
        		\includegraphics[width=\textwidth]{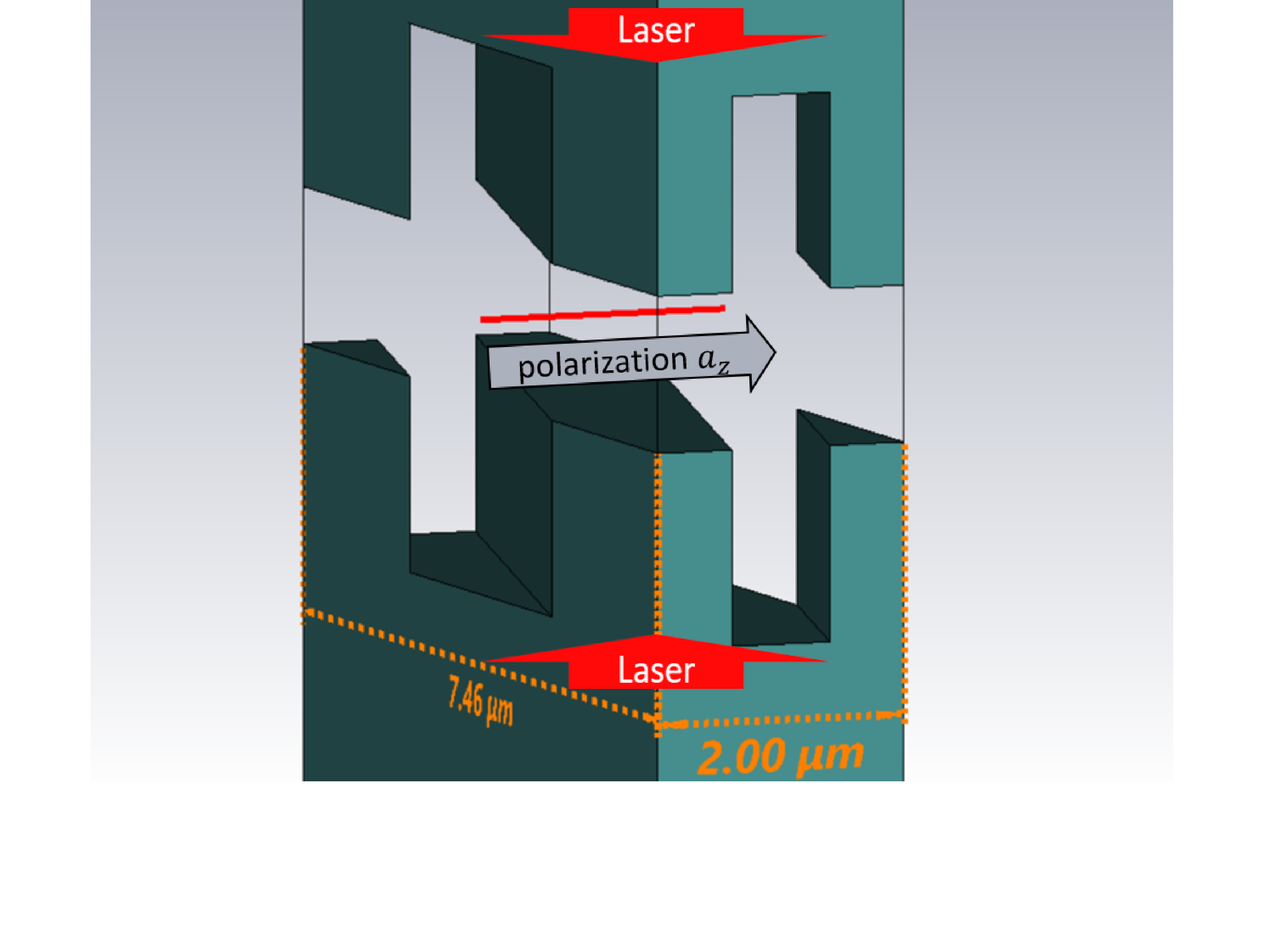}
        		\caption{Geometry of a tilted DLA cell with the electron beam indicated as a red line and a $z$-polarized laser field $a_{\rm{z}}$. The aperture is $\Delta y = 1.2\mu$m and the fill factor $r_{\rm{f}}$ of silica per grating period is 60\%.}
        		\label{sas_fig:undulator_sketch}
        	\end{minipage}
        	\hfill
        	\begin{minipage}{0.49\textwidth}
        	    \vspace{0pt}
        		\includegraphics[width=\textwidth]{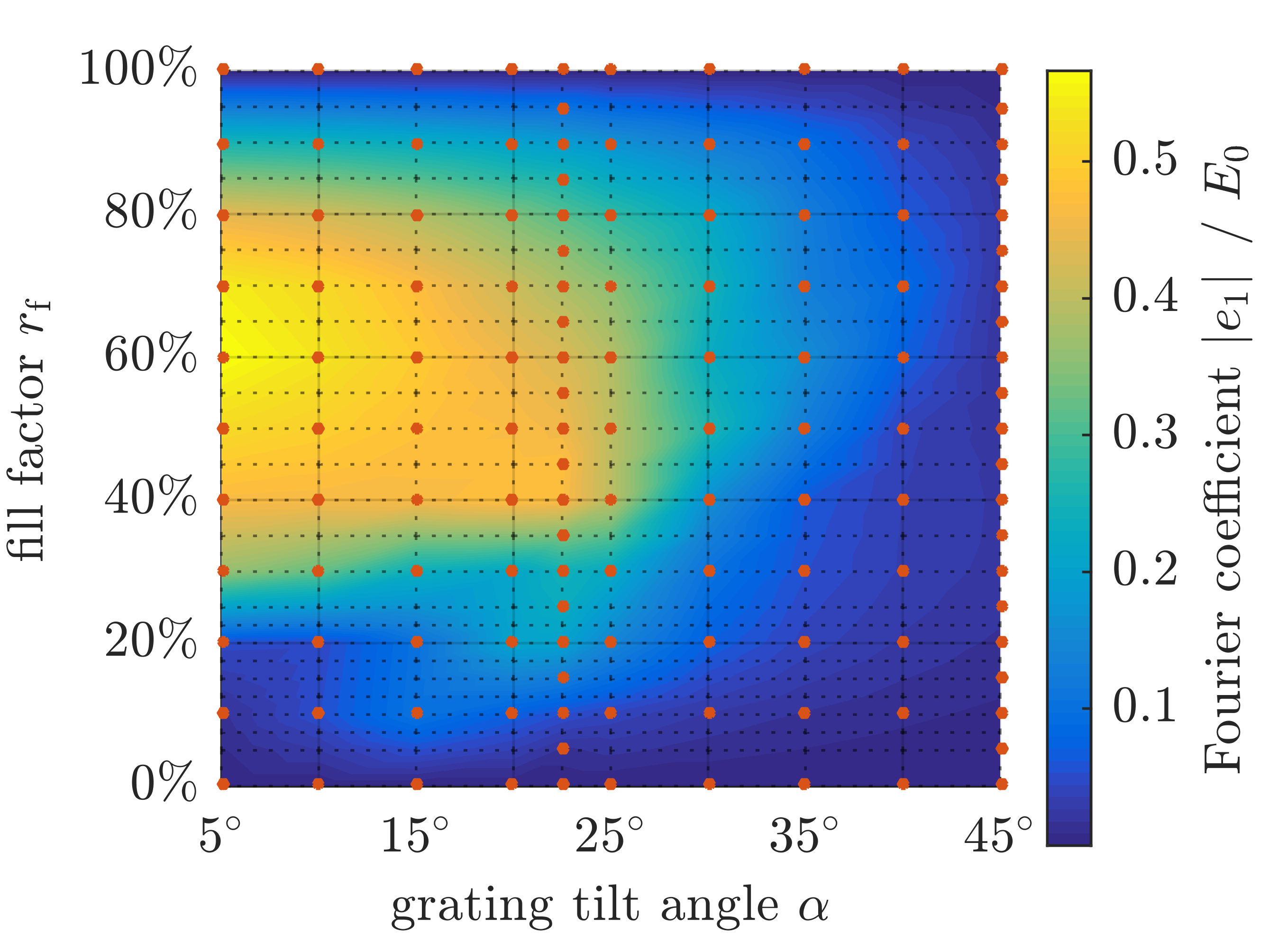}
        		\caption{Fourier coefficient $e_1$ for different fill factors $r_{\rm{f}}$ and various tilt angles $\alpha$. Orange dots indicate simulated geometries, the colorscale in between is interpolated.}
        		\label{sas_fig:undulator_structure_e1}
        	\end{minipage}
        \end{figure}

        Figure\,\ref{sas_fig:undulator_sketch} shows one cell of a tilted DLA structure composed of two opposing silica ($\epsilon_{\rm{r}} = 2.0681$) diffraction gratings for the laser wavelength $\lambda= 2\pi/k = 2\mu$m. The laser excites a grating-periodic electromagnetic field with $k_{\rm{z}} = 2\pi/\lambda_{\rm{g}}$ which imposes a deflection force \cite{Plettner2008Microstructure-basedLaser} on the electrons. Our investigation considers two different concepts for the application of tilted DLA gratings as undulators. First, the concept introduced in refs. \cite{Plettner2008Microstructure-basedLaser, Plettner2008ProposedUndulator} which uses a phase-synchronous DLA structure fulfilling the Wideroe condition Eq. \eqref{Eq:Wideroe} (see ref. \cite{Niedermayer2017BeamScheme} for an analysis of the dynamics therein). Second, a concept similar to microwave \cite{Shintake1982MicrowaveUndulator}, terahertz \cite{Rohrbach2019THz-drivenSource} or laser \cite{Toufexis2014AUndulator} driven undulators which uses a non-synchronous DLA structure that does not fulfill Eq.~\eqref{Eq:Wideroe}. 

        Efficient operation of the DLA undulator requires a design with optimized cell geometry to maximize the interaction of the electron beam with the laser field. Figure\,\ref{sas_fig:undulator_structure_e1} shows simulation results for a parameter scan of the tilt angle $\alpha$ and the fill factor $r_{\rm{f}}$ which is the tooth width divided by the grating period. The tooth height is kept constant at $h=1.5\mu$m. The Fourier coefficient $e_1$ at the aperture center indicated by the red line as defined in Eq.~\eqref{Eq:e1def} is a figure of merit for the interaction strength. For a DLA structure with reasonable aperture $\Delta y = 1.2\mu$m and tilt angle $\alpha \approx 25$ degrees the available structure constant is $\left|e_1\right|/E_0 \approx 0.4$. At 2~$\mu$m, a reasonably short (three digit fs) laser pulse provides at the damage threshold of silica a maximum field strength of $E_0 \approx 1\dots2$~GV/m.

\subsection{Analytical Model for the Non-Synchronous Undulator}
        The investigation of the non-synchronous DLA undulator concept allows to use an analytical model similar to Ref.~\cite{Tran1987Free-ElectronWigglers}. The Hamiltonian for electrons with charge $q$, mass $m_{\rm{0}}$, momentum components $p_i = \gamma \beta_i$, $i=\{x,y,z\}$ in a electromagnetic vector potential $\mathbf{a}\left(x,y,z,ct\right)$ defined with dimensionless quantities is
            \begin{equation}
                H\left(x,p_{\rm{x}},y,p_{\rm{y}},ct,\gamma;z\right) = - \sqrt{\gamma^2 - 1 - \left(p_{\rm{x}} - a_{\rm{x}}\right)^2 - \left(p_{\rm{y}} - a_{\rm{y}}\right)^2} - a_{\rm{z}}~\rm{.}
                \label{sas_eqn:hamilton_dla}
            \end{equation}
        Note that since the independent parameter is $z$, the Hamiltonian represents the longitudinal momentum $p_z$ and not the total energy.    
        The electromagnetic field along the electron beam path of a tilted DLA structure can be written as a periodic vector potential in the form of
            \begin{equation}
                \mathbf{a}\left(x,y,z,ct\right) = a_{\rm{z}} \cosh{\left(k_{\rm{y}}\,y\right)} \sin{\left(k\,ct - k_{\rm{z}}\,z + k_{\rm{x}}\,x\right)} \ \mathbf{e}_{\rm{z}}
                    \label{sas_eqn:vectorpotential_dla}
            \end{equation}
        with the reciprocal grating vectors of the tilted DLA cell $k_{\rm{z}}$ and $k_{\rm{x}} = k_{\rm{z}} \tan{\alpha}$ (see ref. \cite{Niedermayer2017BeamScheme}), $k_{\rm{y}} \equiv \sqrt{\left|{k}^{2} - {k_{\rm{x}}}^{2} - {k_{\rm{z}}}^{2}\right|}$, and the dimensionless amplitude defined as
            \begin{equation}
                a_{\rm{z}} \equiv \frac{q \left|e_1\left(\alpha\right)\right|/k}{m_{\rm{0}} c^2}~\rm{.}
            \end{equation}
            
\begin{figure}
        	\begin{minipage}{0.49\textwidth}
        		\includegraphics[width=\textwidth]{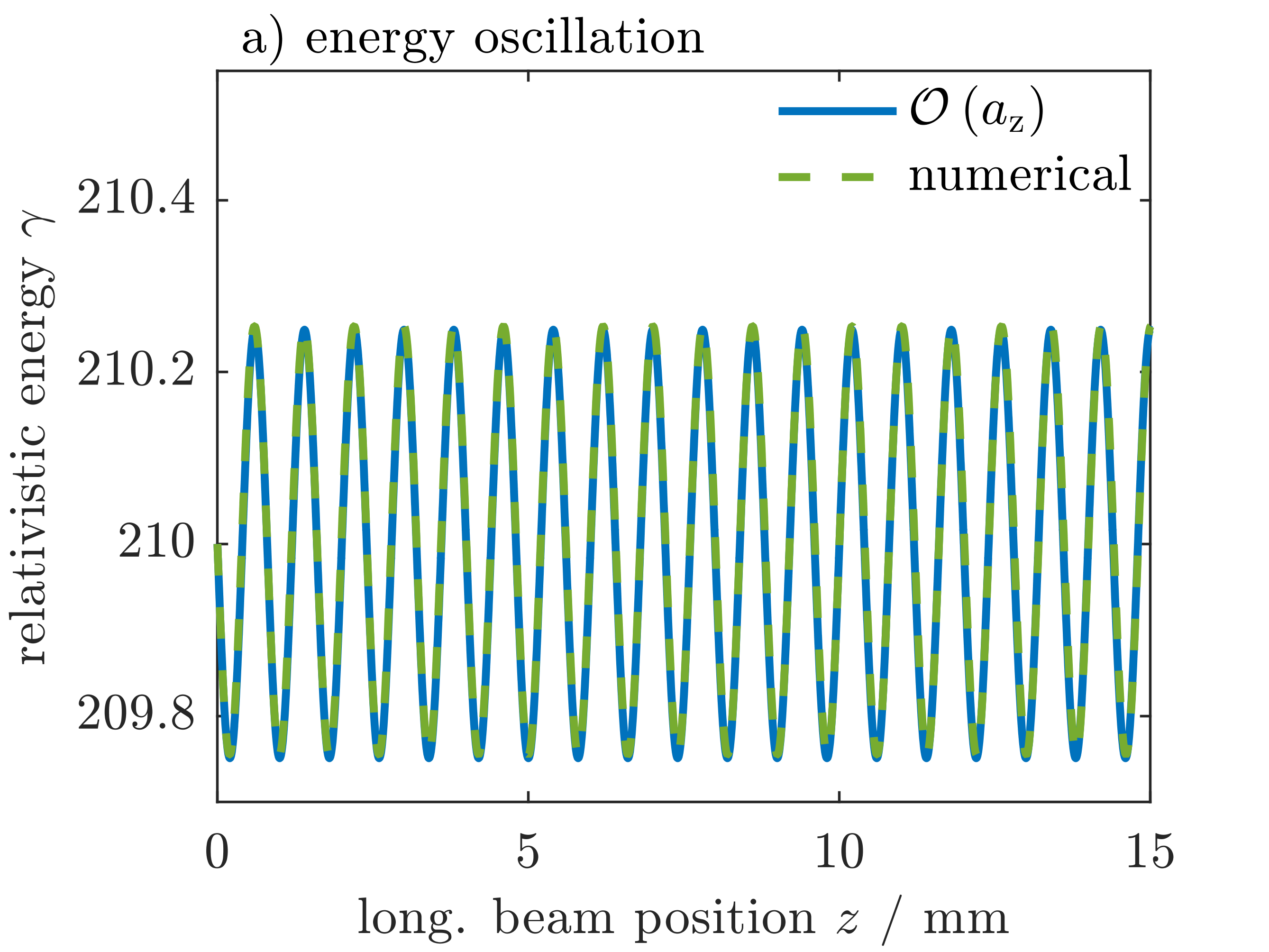}
        	\end{minipage}
        	\hfill
        	\begin{minipage}{0.49\textwidth}
        		\includegraphics[width=\textwidth]{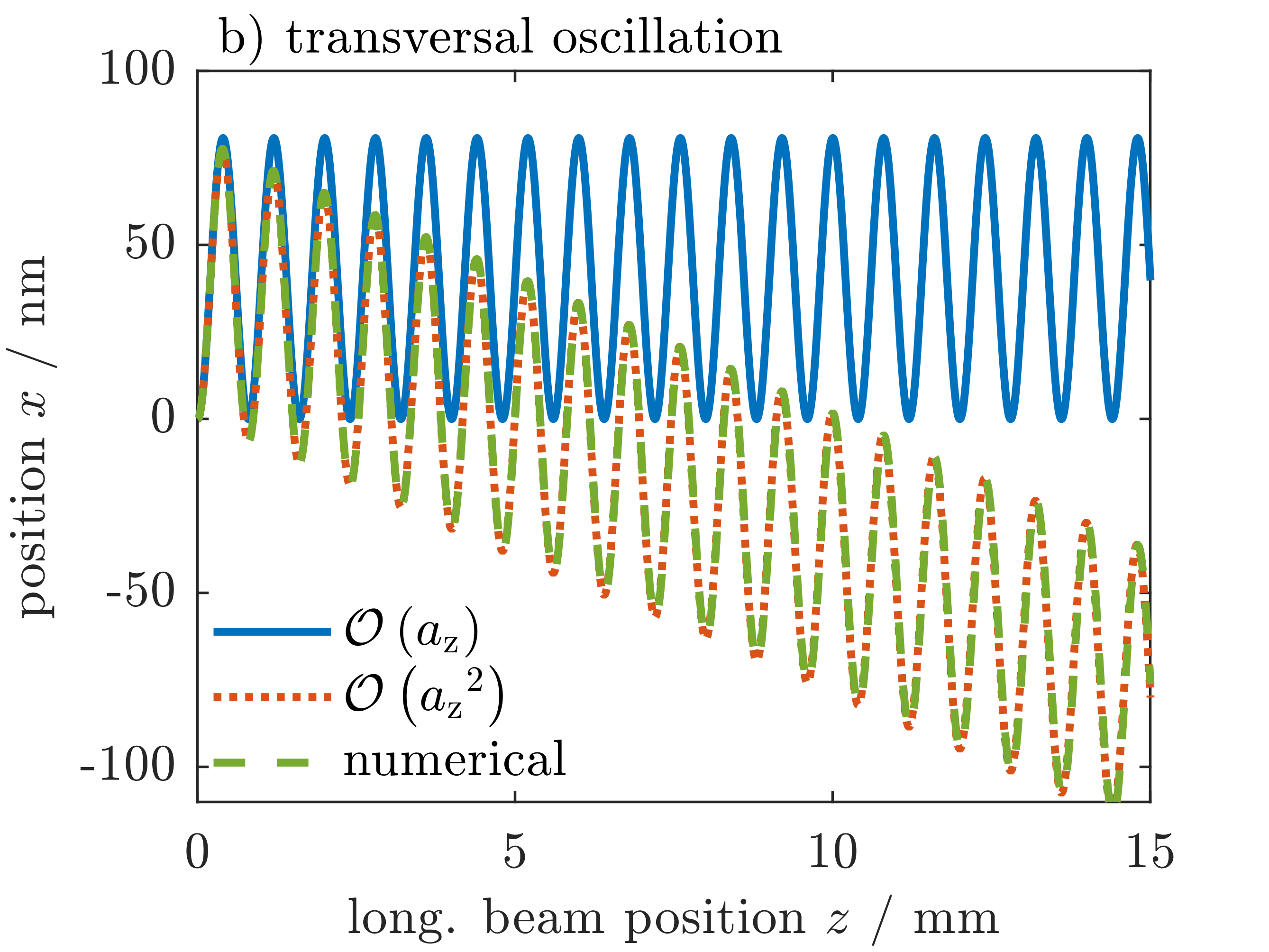}
        	\end{minipage}
        	\caption{Comparison of the $\mathcal{O}\left({a_{\rm{z}}}\right)$ approximations with the numerical reference solution of Eq. \eqref{sas_eqn:hamilton_dla}. The electromagnetic interaction with the laser field induces an oscillation of the a) kinetic energy and b) transversal position. The continuous drift motion shown in b) is a second order $\mathcal{O}\left({a_{\rm{z}}}^2\right)$ effect.}
        	\label{sas_fig:undulator_perturbation_comparison}
        \end{figure}

        In general, Eq.~\eqref{sas_eqn:hamilton_dla} and Hamilton's equations yield six coupled nonlinear differential equations for the phase space coordinates $x$, $p_{\rm{x}}$, $y$, $p_{\rm{y}}$, $ct$, and $\gamma$ as a function of the independent variable $z$. For a DLA undulator with $E_0 \sim 1$~GeV/m electric field strength, the rest mass of an electron is much larger than its energy modulation across one laser wavelength. Consequently, the amplitude of the dimensionless vector potential, $a_{\rm{z}} \approx 6.2e-4 \ll 1$, is small and allows to calculate the solutions of Eq. \eqref{sas_eqn:hamilton_dla} by perturbation. Taking into account the second order terms $\mathcal{O}\left({\gamma_0}^{-2}\right)$ and the first order terms $\mathcal{O}\left({a_{\rm{z}}}\right)$ for the $107$~MeV beam yields analytic approximations for the energy $\gamma\left(z\right)$ and the transverse position $x\left(z\right)$ of the electron. Figure\,\ref{sas_fig:undulator_perturbation_comparison} compares the approximations with numerically computed solutions of Eq.~ \eqref{sas_eqn:hamilton_dla}. In contrast to a magnetostatic undulator the energy in a DLA undulator oscillates, as can be seen in Fig.\,\ref{sas_fig:undulator_perturbation_comparison}\,a). 
        The analytical approximation for $x\left(z\right)$ yields an adequate estimate for the amplitude and periodicity of the transversal particle oscillation in Fig.\,\ref{sas_fig:undulator_perturbation_comparison}\,b). Adding the $\mathcal{O}\left(a_{\rm{z}}^2\right)$ terms also reproduces the drift motion, sufficient to provide a good agreement with the numerical solution. 
        A synchronicity deviation as compared to Eq.~\eqref{Eq:Wideroe} leads to a drift of the drive laser phase with respect to the electron beam. Accordingly, the deflection in $x$-direction alternates its sign, although the tilt angle remains constant~\cite{Niedermayer2018ChallengesAcceleration}. Thus, the wave number $k_{\rm{u}} = 2 \pi / \lambda_{\rm{u}}$ of the DLA undulator can be determined as
        \begin{equation}
                k_{\rm{u}} \approx \frac{1}{\beta} k - k_{\rm{z}} \label{sas_eqn:perturbation_wavenumber}~\rm{.}
        \end{equation}
         The analytical model provides design guidelines for the experimental realization of an DLA undulator. In Eq. \eqref{sas_eqn:perturbation_wavenumber} the deviation of $k$ with respect to a synchronous DLA structure determines the undulator wavelength $\lambda_{\rm{u}}$. Hence, altering the laser frequency allows direct adjustments of $\lambda_{\rm{u}}$. Aside from the oscillatory deflection the longitudinal field $a_{\rm{z}}$ induces a transversal drift motion which depends on the relative phase 
        \begin{equation}
            \varphi_0 \equiv k \, ct_0 + k_{\rm{z}} \tan\alpha \, x_0  \label{sas_eqn:perturbation_phase} ~\rm{.}
        \end{equation}
        In exactly the same way as for a magnetostatic undulator this effect might be mitigated by smoothly tapering the deflection field amplitude towards both undulator ends. For an electron in the center of the beam channel the undulator parameter~\cite{Schmuser2014SpringerRegime} in the analytical model is
         \begin{equation}
             K_{\rm{z}} = a_{\rm{z}} \frac{k_{\rm{x}}}{k_{\rm{u}}} = \frac{q}{m_{\rm{0}} c^2} \frac{k_{\rm{z}}}{k k_{\rm{u}}}  \left|e_1\left(\alpha\right)\right| \tan \alpha ~\rm{.}\label{sas_eqn:undulator_parameter}
         \end{equation}
         
\begin{figure}
        	\begin{minipage}{0.49\textwidth}
        		\includegraphics[width=\textwidth]{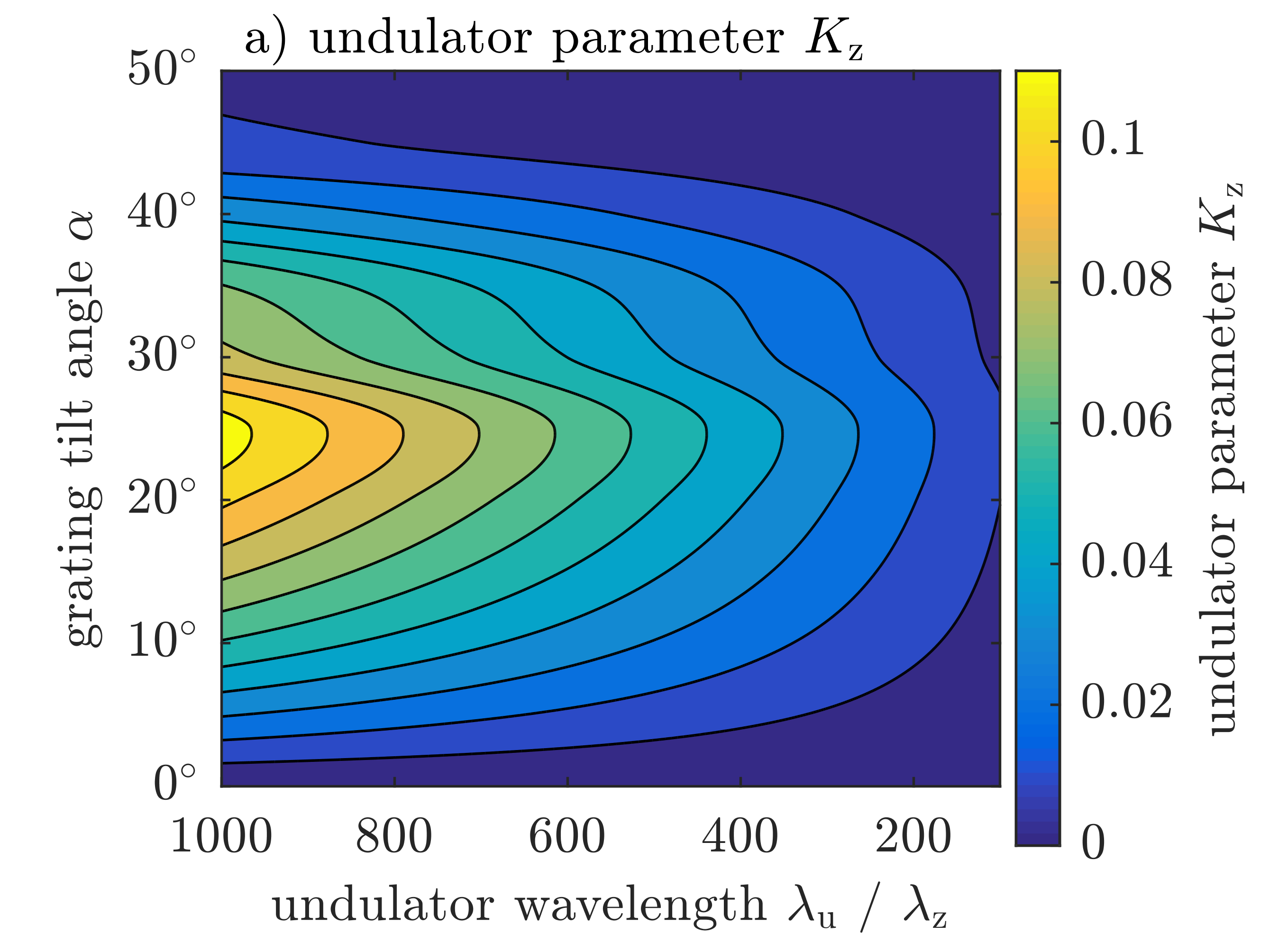}
        	\end{minipage}
        	\hfill
        	\begin{minipage}{0.49\textwidth}
        		\includegraphics[width=\textwidth]{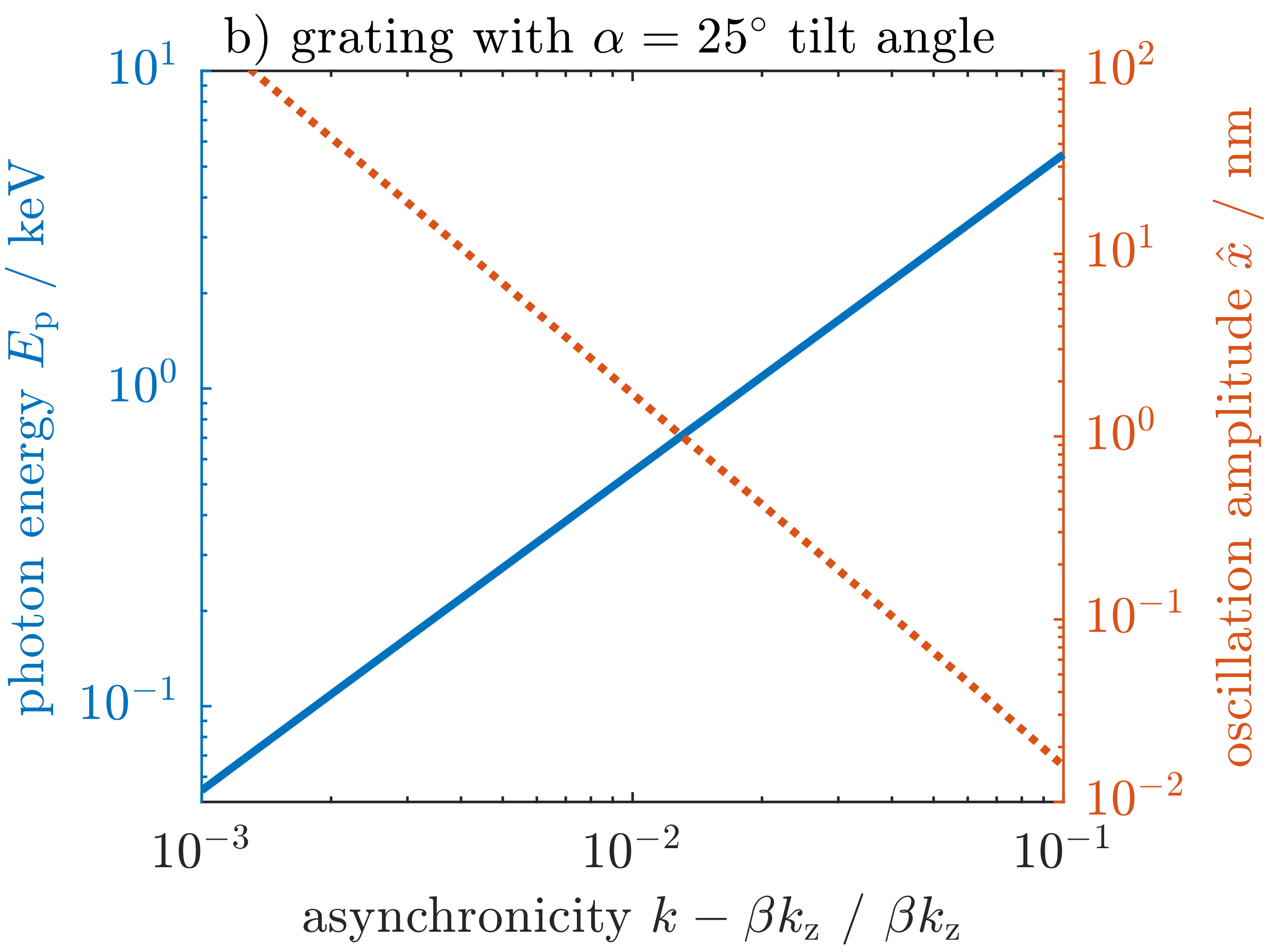}
        	\end{minipage}
        	\caption{The undulator parameter $K_{\rm{z}}$ in a) for the silica DLA grating reaches its maximum at $\alpha \approx 25$ degrees. The photon energy $E_{\rm{p}}$ and the transversal oscillation amplitude $\hat{x}$ in b) depend on the deviation between $k$ and the synchronous DLA wave number $\beta k_{\rm{z}}$ of the laser field.}
        	\label{sas_fig:undulator_properties}
        \end{figure}

        Figure\,\ref{sas_fig:undulator_properties}\,a) shows the dependency of the undulator parameter $K_{\rm{z}}$ on the grating tilt angle $\alpha$ and the undulator wavelength $\lambda_{\rm{u}}$ in a vertically symmetric opposing silica grating structure. For this purpose the synchronous harmonic $e_1$ at the center of the structure was determined as function of the tilt angle. The undulator parameter $K_{\rm{z}}$ shows a local maximum at an tilt angle of $\alpha \approx 25$ degrees. Furthermore, $K_{\rm{z}}$ increases linearly with the undulator wavelength $\lambda_{\rm{u}}$. We investigate a design using $\lambda_{\rm{u}} = 400 \lambda_{\rm{z}}$ which corresponds to an effective undulator parameter of $K_{\rm{z}} \approx 0.045$. In Fig.\,\ref{sas_fig:undulator_properties}\,b) the detuning with respect to the synchronous operation $k - \beta k_{\rm{z}}$ determines the transversal oscillation amplitude $\hat{x}$ and the energy of the generated photons $E_{\rm{p}}$. For 0.25\% deviation from synchronicity, the silica DLA undulator induces a $\hat{x} \approx 30$~nm electron beam oscillation and a wavelength of \cite{Schmuser2014SpringerRegime}
        \begin{equation}
            \lambda_{\rm{p}} = \frac{\lambda_{\rm{u}}}{2 {\gamma_0}^2} \left(1 + \frac{{K_{\rm{z}}}^2}{2}\right) \approx 9~\rm{nm,}
        \end{equation}
        corresponding to soft X-rays with $E_{\rm{p}} = 0.14$~keV.

    \subsection{Simulation of the Beam Dynamics in Tilted Gratings}
    
        Using the particle tracking code DLATrack6D~\cite{Niedermayer2017BeamScheme} we investigate the beam dynamics in both a synchronous as well as a non-synchronous DLA undulator. Each undulator wavelength $\lambda_{\rm{u}} = 800\mu$m of the investigated structure consists of $400$ tilted DLA cells which are joined along the $z$-direction. The total length of the undulator is 16.4~mm which corresponds to $8200$ DLA cells with $\lambda_g = 2\mu$m or $\approx20$ undulator periods. In order to alternate the deflection for an oscillatory electron motion the relative laser phase needs to shift by $2 \pi$ in total as the beam passes one undulator wavelength. For that reason, the synchronous DLA undulator design introduces a $\pi$ phase shift after each $\lambda_{\rm{u}}/2$. In an experimental setup this can be achieved either by drift sections such as used in the APF scheme or by laser pulse shaping e.g. by a liquid crystal phase mask. In the non-synchronous undulator the drift of the drive laser phase with respect to the electron beam automatically introduces the required shift to modulate the deflection force. Hence, subsequent grating cells automatically induce an oscillatory electron motion.

        Synchronous DLA undulators have already been simulated with DLAtrack6D, see~\cite{Niedermayer2017BeamScheme}. 
        If the phase drift per DLA cell is small ($\Delta \varphi = k_{\rm{u}} \lambda_{\rm{z}} \ll 1$), DLAtrack6D can also be applied to non-synchronous structures. Considering for instance the y-momentum kick, linearization of Eq. \eqref{sas_eqn:hamilton_dla} including terms up to $\mathcal{O}\left({\Delta \varphi}^2\right)$ yields 
        \begin{equation}
                    \Delta y^{\prime} = \frac{\Delta p_{\rm{y}}}{\gamma_0 \beta_0} = - k_{\rm{y}} \lambda \frac{a_0}{\gamma_0} \frac{\left|e_1\left(\alpha\right)\right|}{E_0} \sinh{\left(k_{\rm{y}} \, y_0\right)} \sin{\varphi_0} + \mathcal{O}\left(\beta_{\rm{x}}\right)+\mathcal{O}\left(\beta_{\rm{y}}\right)+\mathcal{O}\left(\frac{k_{\rm{u}}}{k_{\rm{z}}}\right) ~\rm{.}\label{sas_eqn:dlatrack6d_yp}
        \end{equation}
        The first term recovers the tracking equation of DLATrack6D (see ref. \cite{Niedermayer2017BeamScheme}). The correction terms contribute less than 1\% for the investigated structures.
        
\begin{figure}
        	\begin{minipage}{0.49\textwidth}
        		\includegraphics[width=\textwidth]{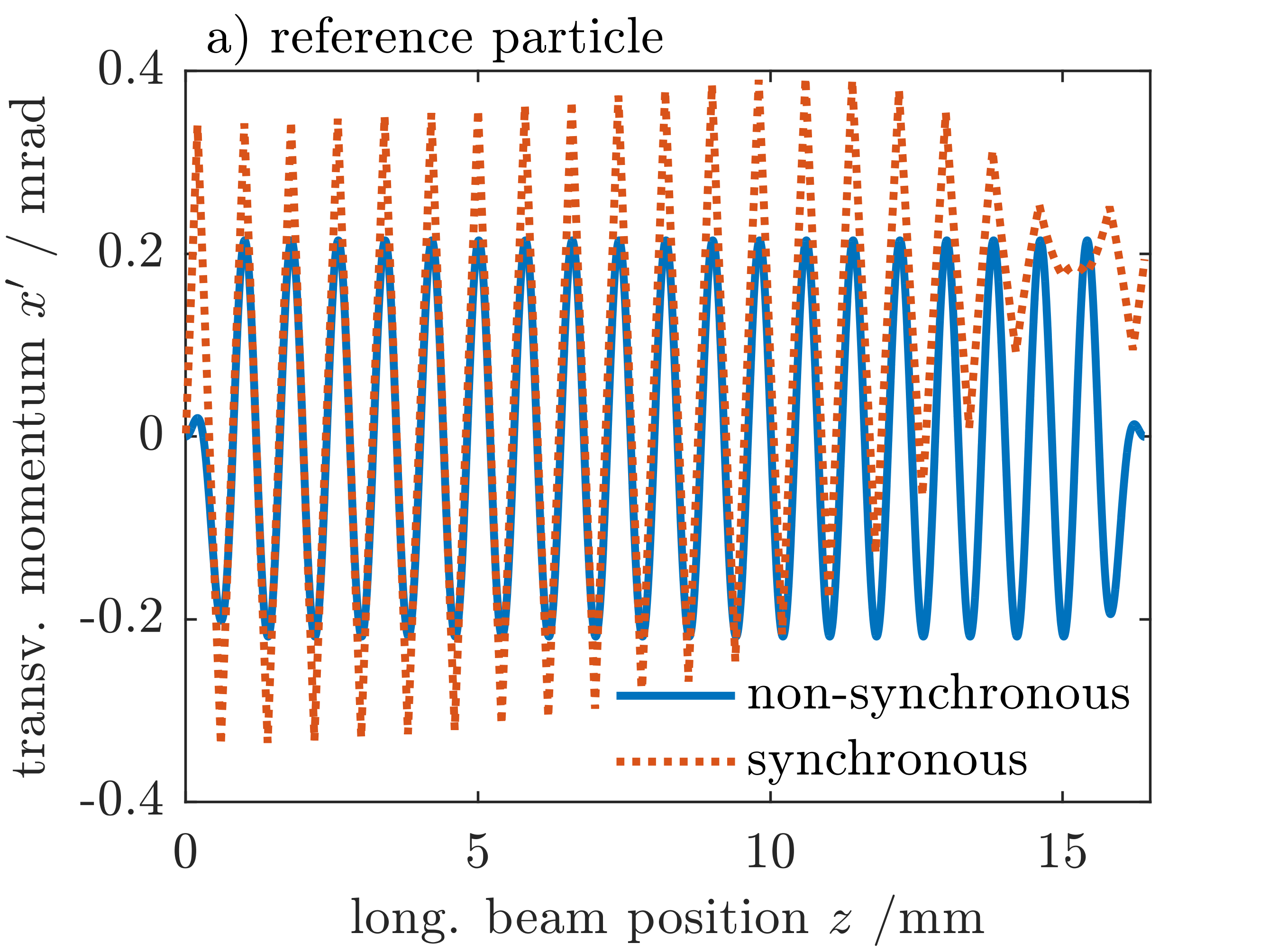}
        	\end{minipage}
        	\hfill
        	\begin{minipage}{0.49\textwidth}
        		\includegraphics[width=\textwidth]{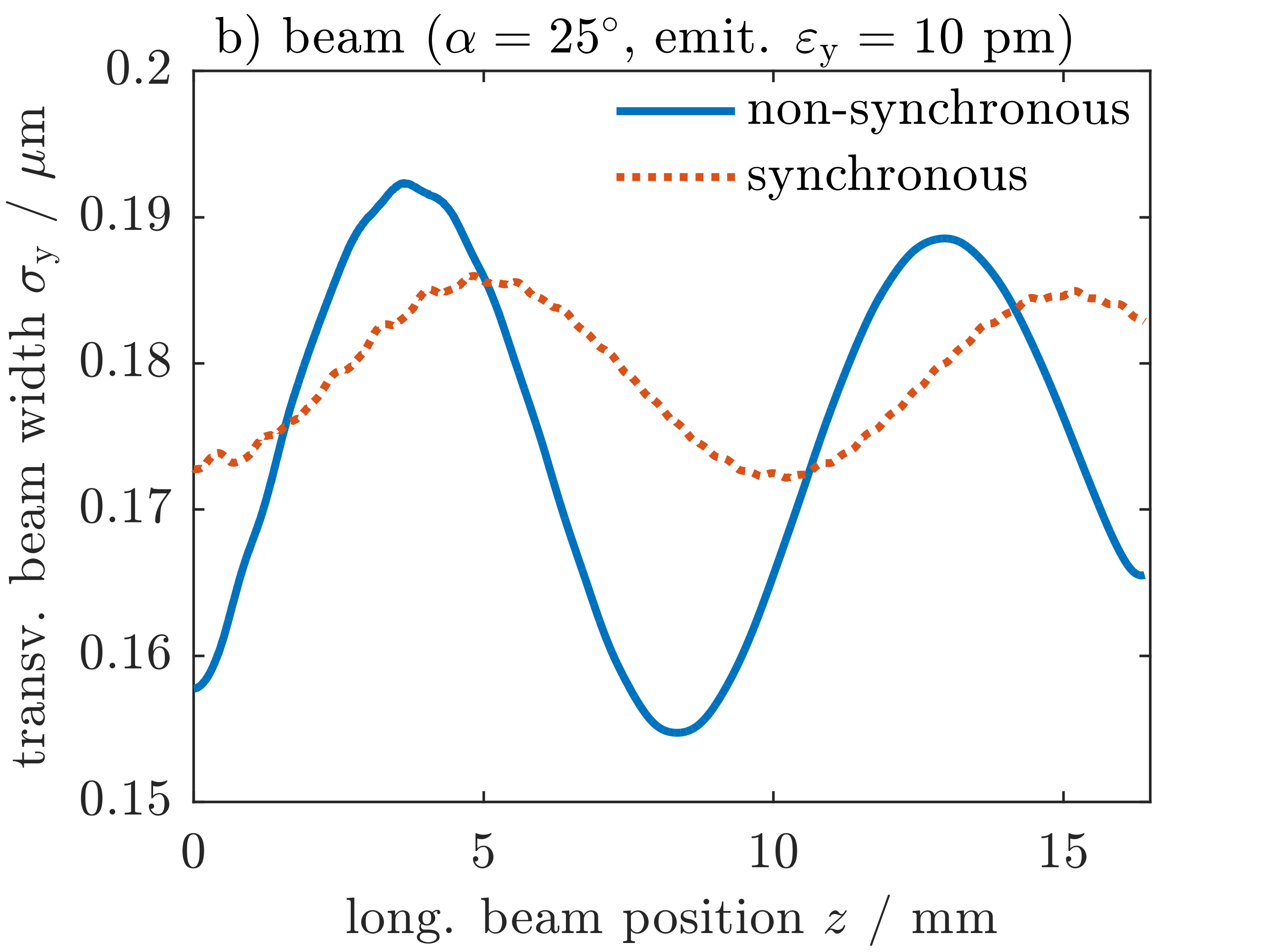}
        	\end{minipage}
        	\caption{The variation of the $x$-coordinate in a) for the reference particle in an non-synchronous DLA undulator with $\lambda_{\rm{u}} = 0.8$mm is sinusoidal. In the synchronous structure the abrupt phase shifts $\Delta \varphi = \pi$ after each $\lambda_{\rm{u}} / 2$ introduce contributions of higher harmonics. Regarding the y-coordinate in b), both structures show an oscillating beam width.}
        	\label{sas_fig:undulator_tracking_reference}
        \end{figure}

        Figure\,\ref{sas_fig:undulator_tracking_reference}\,a) compares tracking results for the particle at the beam center of a synchronous and a non-synchronous DLA undulator. In the synchronous undulator the phase jumps by $\Delta \varphi_0 = \pi$ due to a $\lambda_g/2$ drift section after each $\lambda_{\rm{u}} / 2$ such that deflection force acting on the particle always switches between its maximum and minimum value. 
        Hence, the reference particle's momentum $x^{\prime}$ changes linearly between two segments. The subsequent triangular trajectory introduces contributions of higher harmonics into the radiation. 
        Furthermore, the accumulation of deflections leads to a deviation from the reference trajectory for $z \geq 10$~mm and the accumulated extra distance travelled by the reference particle leads to dephasing, which damps the momentum oscillation. 
        In the non-synchronous DLA undulator the particle trajectory follows a harmonic motion. A smooth phase shift of $\Delta \varphi_0 = 2\pi \lambda_{\rm{z}}/\lambda_{\rm{u}}$ per DLA cell generates a harmonically oscillating deflection which is approximately 30\% smaller compared to the synchronous DLA. A tapering of the deflection strength introduced towards $z = 0$ and $z = 16.4$mm ensures a smooth transition at the ends of the non-synchronous DLA undulator.
        
        Figure\,\ref{sas_fig:undulator_tracking_reference}\,b) shows the width $\sigma_{\rm{y}}$ for an electron beam passing the DLA undulator without particle losses. A transversal geometric emittance of $\varepsilon_{\rm{y}} = 10$~pm ensures 100\% transmission. The simulations use an electron beam with the twiss parameters $\hat{\alpha} = 0$ and $\gamma = 1/\hat{\beta}$ at $z=0$. Depending on the phase $\varphi_0$ the transversal momentum kick \eqref{sas_eqn:dlatrack6d_yp} in a DLA cell can be either focusing or defocusing in y-direction. Hence, the beam width oscillates but remains bounded for both DLA undulators. In order to achieve proper beam matching into the structure a future design study will address the focusing properties of both DLA undulator concepts in more detail. 
        
\begin{figure}
        	\begin{minipage}{\textwidth}
        		\includegraphics[width=\textwidth]{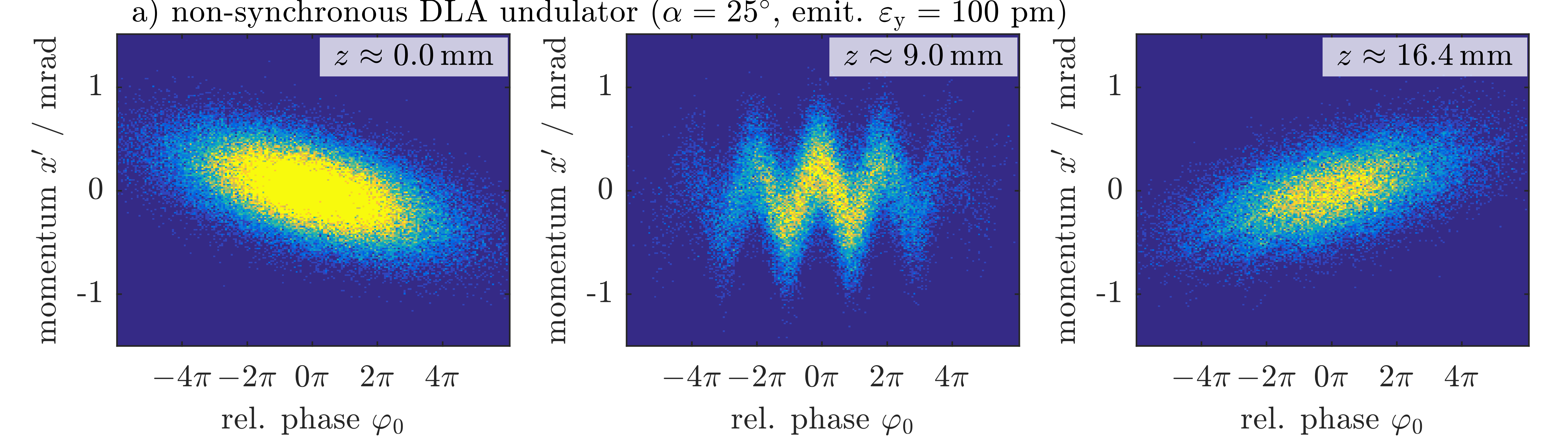}
        	\end{minipage}
        	\hfill
        	\begin{minipage}{\textwidth}
        		\vspace*{11pt}
        		\includegraphics[width=\textwidth]{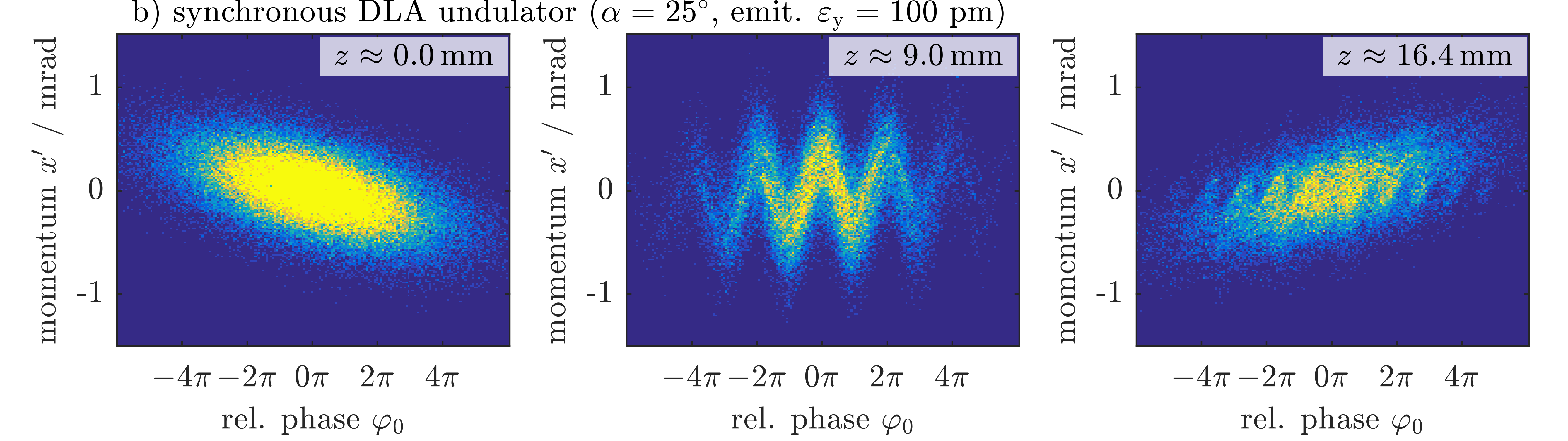}
        	\end{minipage}
        	\caption{Electron beam phase space at the entrance $z \approx 0$mm, in the center $z \approx 9$~mm, and at the exit $z \approx 16$~mm of the a) non-synchronous and b) synchronous DLA undulator structure. Both grating structures reach approximately 50\% transmission of the total beam charge. Note that the horizontal axis is given by the particle phase as defined in Eq.~\ref{sas_eqn:perturbation_phase}, which basically projects the 6D phase space on a diagonal 2D plane.}
        	\label{sas_fig:undulator_tracking_beam}
        \end{figure}

        Figure\,\ref{sas_fig:undulator_tracking_beam} shows the phase space of an electron beam passing a DLA undulator in a) non-synchronous and b) synchronous operation mode. The transversal geometric emittance $\varepsilon_{\rm{x}} = 1$nm and the energy spread $\sigma_{\rm{E}} = 0.02\%$ follow the design parameters of ARES \cite{Panofski2021CommissioningSinbad-ares}. The bunch length is $\sigma_t=1$~fs. The phase space in the center of the undulator at $z \approx 9$~mm shows that both DLA designs induce transversal electron oscillations across the whole beam. However, the transverse electron beam size is larger than one unit cell of the DLA undulator such that the particle distribution transversely ranges across several grating periods. For this reason the momentum $x^{\prime}$ at $z \approx 9$~mm varies depending on the relative phase $\varphi_0$ in Eq. \eqref{sas_eqn:perturbation_phase} of the electron with respect to the laser field. The averaged momentum of the particle beam remains zero. In the non-synchronous operation mode the particles experience an averaged deflection and focusing force. Thus, all electron trajectories are similar, but differ by a constant drift motion along the x-coordinate. The drift depends on the initial phase $\varphi_0$ at which the particle enters the undulator. In the synchronous mode each particle experiences a different deflection and focusing force which accumulates additive along one undulator period $\lambda_{\rm{u}}$. The oscillation of each particle depends on its phase $\varphi_0$. Thus, a substructure in the phase space slightly visible at $z \approx 9$mm and more prominent towards the exit at $z \approx 16.4$~mm develops as the beam passes the undulator.

\section{Conclusion}
For the study of beam dynamics in DLA, computer simulations will remain essential. With combined numerical and experimental approaches, the challenges of higher initial brightness and brightness preservation along the beamline can be tackled. 

The electron sources available from electron microscopy technology are feasible for experiments, however cost and size puts a major constraint on them. The upcoming immersion lens nanotip sources offer a suitable alternative. Their performance does not reach the one of the commercial microscopes yet, but one can expect significant improvements in the near future. This will enable low energy DLA experiments with high energy gain and full six-dimensional confinement soon. 

Full confinement is a requirement for high energy gain at low injection energies, since the low energy electrons are highly dynamical. Recent APF DLA experiments showed that (as theoretically expected) the so-called invariant dimension is in fact not invariant for the mostly used silicon pillar structures. The consequences are energy spread and emittance increase, eventually leading to beam losses. A way to overcome this is to turn towards a 3D APF scheme, which can be implemented on commercial SOI wafers. The 3D scheme has also advantages at high energy, since it avoids the focusing constants going to zero in the ultrarelativistic limit. Only the square-sum goes to zero and thus a counterphase scheme is possible with high individual focusing constants. Using a single high damage threshold material for these structures leads however to fabrication challenges.

The spatial harmonic focusing scheme is much less efficient than APF, since most of the damage threshold limited laser power goes into focusing rather than into acceleration gradient. However, when equipped with a focusing scheme imprinted on the laser pulse by a liquid crystal phase mask, it can operate on a generic, strictly periodic grating structure. This provides significantly improved experimental flexibility. Moreover, as the scheme intrinsically operates with different phase velocities of electromagnetic waves in the beam channel, it can be easily adapted to travelling wave structures. 

For a high energy collider, travelling wave structures are definitely required to meet the laser energy efficiency requirement. They can efficiently transfer energy from a co-propagating laser pulse to the electrons, until the laser pulse is depleted. Laterally driven standing wave structures cannot deplete the pulse. In the best case, on can recycle the pulse in an integrated laser cavity~\cite{Siemann2004EnergyAccelerators}. However, significant improvement in energy efficiency as compared to the status quo can be obtained by waveguide driven DLAs, see~\cite{Hughes2018On-ChipAccelerators,Sapra2020On-chipAccelerator}. More information about the requirements and the feasibility of DLA for a high energy collider can be found in~\cite{England2022ConsiderationsAccelerators}.

The on-chip light source is still under theoretical development. Currently we outline a computationally optimized silica grating geometry as well as an analytical description and numerical simulations of the dynamics for electrons passing a soft X-ray radiation DLA undulator. The analytical model provides essential guidelines for the ongoing design process. The concept of a non-synchronous tilted grating structure turns out to be a promising alternative to the synchronous operation mode. The non-synchronous undulator operates without phase jumps in the structure, which relaxes the fabrication requirements and the requirements on the drive laser phase front flatness. Furthermore, variation of the laser wavelength allows direct fine tuning of the undulator period length. Preliminary results indicate that in order to achieve approximately 50\% beam transmission, the geometric emittance must not exceed $\varepsilon_{\rm{y}} = 100$~pm (at 107~MeV). Optimization of the beam focusing within the DLA undulator structures is outlined for investigations in the near future.

\section*{Acknowledgements}

We acknowledge funding by the Gordon and Betty Moore Foundation under grant GMBF4744 (ACHIP) and the German Ministry of Education and Research (Grant No. 05K19RDE) as well as funding from the state of Hessen via LOEWE Exploration.

\bibliographystyle{unsrtnat}
\bibliography{Mendeley}

\end{document}